\documentclass[10pt]{article}

\usepackage[english]{babel} 
\usepackage{amsmath}                
\usepackage{amsfonts}               
\usepackage{amssymb}                
\usepackage{amsopn}                 
\allowdisplaybreaks[3] 
\usepackage{amsthm}                
\usepackage{bbm}                    
\usepackage{mathrsfs}               
\usepackage[retainorgcmds]{IEEEtrantools} 
\usepackage{calc}                   
\usepackage[dvips]{graphicx}        
\usepackage{epsfig}                
\usepackage{psfrag}                 
\usepackage[dvips]{color}           
\usepackage{fancyhdr}               
\usepackage{verbatim}               
\usepackage{exscale}                
\usepackage{macros}
\newtheorem{theorem}{Theorem}
\newtheorem{lemma}[theorem]{Lemma}

\newtheorem{proposition}[theorem]{Proposition}

\newtheorem{remark}{Remark}
\newtheorem{example}{Example}

\newcommand{\Cfb}{\dot{C}_{\textnormal{FB}}(0)}
\newcommand{\CFB}{C_{\textnormal{FB}}(\SNR)}
\renewcommand{\Var}[1]{\textnormal{\textsf{Var}}\!\left({#1}\right)} 

\title{Channels that Heat Up}

\author{Tobias Koch ~~ Amos Lapidoth\\\small ETH Zurich\\\small Zurich, Switzerland\\
   \small Email: \{tkoch, lapidoth\}@isi.ee.ethz.ch
 \and 
 Paul P.~Sotiriadis\\\small Johns Hopkins University\\\small Baltimore, MD, USA\\\small Email: pps@jhu.edu
 }

\date{}

\begin{document}

\maketitle

\begin{abstract}
  \renewcommand{\thefootnote}{}
  This work considers an additive noise channel where the time-$k$
  noise variance is a weighted sum of the channel
  input powers prior to time $k$. This channel is motivated by
  point-to-point communication between two terminals that are embedded
  in the same chip. Transmission heats up the entire chip and hence increases
  the thermal noise at the receiver. The capacity of this channel (both with
  and without feedback) is studied at low transmit powers and at high
  transmit powers.

  At low transmit powers, the slope of the capacity-vs-power
  curve at zero is computed and it is shown that the heating-up effect is
  beneficial. At high transmit powers, conditions are determined under
  which the capacity is bounded, i.e., under which the capacity does not
  grow to infinity as the allowed average power tends to infinity.
  \footnote{The material in this paper
  was presented in part at the 2007 IEEE International Symposium on Information
  Theory (ISIT), Nice, France, and at the 2007 IEEE Information Theory
  Workshop (ITW), Lake Tahoe, CA, USA.}
\end{abstract}
\setcounter{footnote}{0}

\section{Introduction}
\label{sec:intro}
Thermal heating in electronic systems is strongly related to
performance limitation, aging, reliability and safety issues.
High performance-density and small physical size (area or volume)
make thermal heating important and challenging to address.
This is enhanced by the trend of modern (micro-)electronics
technology to pack more and faster operations within the smallest
possible physical area in order to increase performance, reduce
cost and size, and therefore expand the potential applications of
the product and make it more profitable.

Electrical power dissipation into heat raises the local
temperature of the circuit; more accurately, the temperature
depends on the circuit activity. The temperature influences the
power of the intrinsic noise in the circuit which in turn
reduces the effective communication or computation capacity of the
circuit. This ``negative'' performance feedback is expected to
become a bottleneck of future technology
\cite{venkatesan01}, \cite{kish02}.

This work aims to add this dimension to our understanding of the
coupling mechanism between communication and computation
performance and thermal heating.
To this end a class of communication channels is introduced,
where the channel's noise power depends dynamically on the
channel's activity, and its channel capacity is studied.


To support the previous statements and motivate the mathematical
development of this new class of channels we first discuss the
underlying physical mechanism that connects circuit activity with
power consumption and thermal heating.
%
%
Thermal heating is unavoidable in electronic circuits. Every
circuit block converts part of the power it draws from the power
supply network (and to certain extent from its interconnections
with other blocks) into heat which raises the local temperature.

A circuit block in a microchip occupies certain physical space
within which heat is distributively generated and diffused
according to the \emph{heat diffusion equation} (ignoring other heat
sources)
\begin{eqnarray}\label{eq-distr-diffusion}
 \textsf{C}_{\textnormal{hv}}\frac{\partial T}{\partial t} = \nabla \cdot\left(\frac{1}{\rho_{\textnormal{thd}}}\nabla T\right) +
\const{E}'
\end{eqnarray}
where $\textsf{C}_{\textnormal{hv}}$ is the volumetric heat capacity
of the material, $\partial T/\partial t$ is the change in temperature
over time, $\nabla\cdot$ is the divergence, $\rho_{\textnormal{thd}}$
is the distributed thermal resistance, $\nabla T$ is the temperature
gradient, and $\const{E}'$ is the power density of the added heat,
\cite{goodson99}, \cite{lienhard08}.

In many cases the diffusion equation can be replaced by the
corresponding \emph{ordinary differential equation} (ODE) that
provides a lumped model of the thermal dynamics.
Consider for example a microchip (die), made out of material of
lower thermal resistance, which is internally heated by the
activity of circuits and transfers the heat to the environment
(e.g., air) which has much higher resistance. In this case we can write
\begin{eqnarray}\label{eq-lumped-diffusion}
\textsf{C}_{\textnormal{h}} \frac{dT}{dt} = \frac{ T_{\textnormal{e}} - T }{\rho_{\textnormal{th}}} + \const{E}
\end{eqnarray}
where $\textsf{C}_{\textnormal{h}}$ is the heat capacity of the microchip (die), $\rho_{\textnormal{th}}$ is
the thermal resistance between the die and the environment (e.g.,
air), $T_{\textnormal{e}}$ is the temperature of the environment, and $\const{E}$ is the
instantaneous heat generated, i.e., the electrical power converted
into heat by the circuit.

Solving \eqref{eq-lumped-diffusion} with the assumption that at time
$t=0$ we have $T=T_e$ with $T_{\textnormal{e}}$ being fixed, we obtain
\begin{eqnarray}\label{eq-lumped-diffusion-sol}
T(t) = T_{\textnormal{e}} + \frac{1}{\textsf{C}_{\textnormal{h}}} \int_{0}^{t}
e^{\frac{\xi-t}{\rho_{\textnormal{th}}\textsf{C}_{\textnormal{h}}}}\const{E}(\xi)d\xi, \qquad t\in\Reals.
\end{eqnarray}
If the circuit operates based on a reference clock of period
$\tau$, \eqref{eq-lumped-diffusion-sol} can be
approximated by its discrete version
\begin{eqnarray}\label{eq-lumped-diffusion-sol-discr}
T_k = T_{\textnormal{e}} + \sum_{\ell=1}^{k-1} \frac{\tau}{\textsf{C}_{\textnormal{h}}}
e^{-\frac{\tau}{\rho_{\textnormal{th}\textsf{C}_{\textnormal{h}}}}(k-\ell)}
\const{E}_{\ell}, \qquad k\in\Integers^+,
\end{eqnarray}
where $\Integers^+$ denotes the set of positive integers, and where
the sequences ${\left\{ T_k \right\}}$ and
${\left\{ \const{E}_k \right\}}$ are the samples at integer multiples of
$\tau$ of $T(\cdot)$ and $\const{E}(\cdot)$, respectively. Equation
\eqref{eq-lumped-diffusion-sol-discr} shows the fading memory
effect of temperature. Note that
\eqref{eq-lumped-diffusion-sol-discr} also captures discrete
versions of distributed or higher order lumped approximations of
the diffusion equation \eqref{eq-distr-diffusion}.

Every electronic circuit has some intrinsically generated noise. This
noise is added to the received signal degrading its quality.
Especially in the popular class of circuits based on MOS
transistors \cite{tsividis03}, this noise is dominated by a thermal
noise component that is stationary Gaussian, and in most
applications it can be considered white. The variance of the
thermal noise $\const{N}$ follows the Johnson-Nyquist formula
\begin{eqnarray}\label{eq-thermal-noise}
\const{N} = \lambda T \WW
\end{eqnarray}
where $\WW$ is the considered bandwidth, $T$ is the temperature of
the receiver circuit block, and $\lambda$ is a proportionality
constant \cite{tsividis03}, \cite{enzcheng00}, \cite{razavi99}.


The transmission of information is typically associated with dissipation of
energy into heat. Thus, in view of
\eqref{eq-lumped-diffusion-sol-discr} and \eqref{eq-thermal-noise}, this motivates a channel
model where the variance $\theta^2$ of the additive noise is determined by the
history of the power of the transmitted signal, i.e.,
\begin{equation}
  \theta^2(x_1,\ldots,x_{k-1}) = \sigma^2 + \sum_{\ell=1}^{k-1}
  \alpha_{k-\ell}x_{\ell}^2, \qquad k\in\Integers^+,
\end{equation}
where $x_{\ell}$ is the transmitted symbol at time
$\ell\in\Integers^+$, and where $\sigma^2$ and $\{\alpha_{\ell}\}$ will be defined
in Section~\ref{sec:channelmodel}.

The rest of this paper is organized as follows. Section~\ref{sec:channelmodel}
describes the channel model in more detail. Section~\ref{sec:capacity}
discusses channel capacity and lists some important properties
thereof. The main results are presented in Section~\ref{sec:result}. The
proofs of the results are given in Sections~\ref{sec:lowSNR} and
\ref{sec:highSNR}. Section~\ref{sec:summary} concludes with a
summary.

\section{Channel Model}
\label{sec:channelmodel}
\begin{figure}
 \centering
 \psfrag{T}[cc][cc]{Transmitter}
 \psfrag{C}[cc][cc]{Channel}
 \psfrag{R}[cc][cc]{Receiver}
 \psfrag{D}[cc][cc]{Delay}
 \psfrag{M}[b][b]{$M$}
 \psfrag{Mh}[b][b]{$\hat{M}$}
 \psfrag{X}[b][b]{$X_k$}
 \psfrag{Y}[b][b]{$Y_k$}
 \psfrag{Yr}[b][b]{$Y_1^{k-1}$}
 \epsfig{file=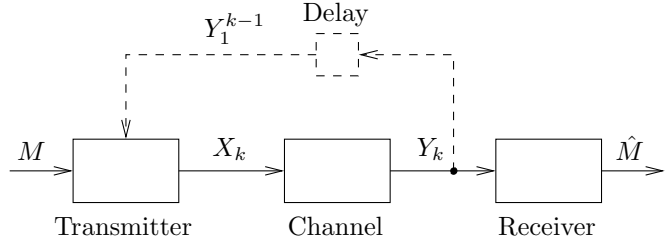, width=0.6\textwidth}
 \caption{A schema of the communication system.}
 \label{fig1}
\end{figure}

We consider the communication system depicted in
Figure~\ref{fig1}. The message $M$ to be transmitted over the channel
is assumed to be uniformly distributed over the set
$\set{M}=\{1,\ldots,|\set{M}|\}$ for some positive integer
$|\set{M}|$. The encoder maps the message to the length-$n$ sequence
$X_1,\ldots,X_n$, where $n$ is the \emph{block-length}. In the absence
of feedback, the sequence $X_1^n$ is a function of the message
$M$, i.e., $X_1^n=\phi_n(M)$ for some mapping $\phi_n:
\set{M} \to \Reals^n$. Here $A_m^n$ stands for $A_m,\ldots,A_n$, and
$\Reals$ denotes the set of real numbers.
If there is a feedback link, then $X_k$, $k=1,\ldots,n$ is not only a function of
the message $M$ but also of the past channel output symbols
$Y_1^{k-1}$, i.e., $X_k=\varphi_n^{(k)}(M,Y_1^{k-1})$ for some mapping
$\varphi_n^{(k)}: \set{M} \times \Reals^{k-1} \to \Reals$.
The receiver guesses the
transmitted message $M$ based on the $n$ channel output symbols
$Y_1^n$, i.e., $\hat{M}=\psi_n(Y_1^n)$ for some mapping $\psi_n: \Reals^n \to \set{M}$.

Conditional on $X_1=x_1,\ldots,X_k=x_k\in\Reals$, the time-$k$ channel output $Y_k \in \Reals$ is given
by
\begin{equation}
  Y_k = x_k +
  \sqrt{\left(\sigma^2+\sum_{\ell=1}^{k-1}\alpha_{k-\ell}x_{\ell}^2\right)}\cdot
  U_k,\qquad k\in\Integers^+,\label{eq:channel}
\end{equation}
where $\{U_k\}$ is a zero-mean, unit-variance, stationary \&
weakly-mixing random process, drawn independently of $M$, and being of
finite fourth moment and of finite differential entropy rate, i.e.,
\begin{equation}
  \E{U_k^4} < \infty \qquad \textnormal{and} \qquad h\big(U_k\big|U_{-\infty}^{k-1}\big) > -\infty.\label{eq:finiteentropy}
\end{equation}
See \cite{petersen83} for a definition of weak mixing.
For example, $\{U_k\}$ could be a stationary \& ergodic Gaussian
process \cite{maruyama49}. In particular, the case of most interest is
when $\{U_k\}$ are independent and identically distributed (IID),
zero-mean, unit-variance Gaussian random variables, and the reader is
encouraged to focus on this case.

The parameter $\sigma^2$ is assumed to be positive. It accounts for
the temperature of the device when the transmitter is silent.
The coefficients $\alpha_{\ell}$, $\ell\in\Integers^+$ are nonnegative
and bounded, i.e.,
\begin{equation}
  \alpha_{\ell} \geq 0, \quad \ell\in\Integers^+ \qquad
  \textnormal{and} \qquad \sup_{\ell \in \Integers^+} \alpha_{\ell} < \infty.\label{eq:bounded}
\end{equation}
They characterize the dissipation of the heat produced by the
transmission of the message $M$.\footnote{It seems reasonable to
  assume that the sequence $\{\alpha_{\ell}\}$ is monotonically nonincreasing, i.e.,
  $\alpha_{\ell} \leq \alpha_{\ell'}$ for $\ell \geq \ell'$. This
  assumption is, however, not required for the results stated in this paper.} 

An example for a heat dissipation profile that satisfies
\eqref{eq:bounded} is the \emph{geometric} heat dissipation profile
where $\{\alpha_{\ell}\}$ is a geometric sequence, i.e.,
\begin{equation}
  \alpha_{\ell} = \rho^{\ell}, \qquad \ell \in\Integers^+ \label{eq:geometric}
\end{equation}
for some $0<\rho<1$. 

The heat dissipation depends \emph{inter alia} on the efficiency of the heat
sink that is employed in order to absorb the produced heat. In the
above example \eqref{eq:geometric}, the heat
sink's efficiency is described by the parameter $\rho$: the smaller
$\rho$, the more efficient the heat sink. In general, an efficient
heat sink is modeled by a heat dissipation profile for which the
sequence $\{\alpha_{\ell}\}$ decays fast.

We study the above channel under an average-power
constraint on the inputs, i.e., the mappings $\phi_n$ (without feedback) and
$\varphi_n^{(1)},\ldots,\varphi_n^{(n)}$ (with feedback) are chosen such that---averaged
over the message $M$ and channel outputs $Y_1^n$---the sequence $X_1^n$ satisfies
\begin{equation}
  \frac{1}{n} \sum_{k=1}^n \E{X_k^2} \leq \const{P}, \label{eq:power}
\end{equation}
and we define the signal-to-noise ratio (SNR) as 
\begin{equation}
  \SNR \triangleq \frac{\const{P}}{\sigma^2}.
\end{equation}

\begin{remark}
The results presented in this paper do not change when \eqref{eq:power}
is replaced by a
\emph{per-message} average-power constraint, i.e., when the mappings $\phi_n$ and
$\varphi_n^{(1)},\ldots,\varphi_n^{(n)}$ are chosen such that, for each
message $m\in\set{M}$ and for any given sequence of output symbols
$Y_1^n=y_1^n$, the sequence $x_1^n$ satisfies
\begin{equation}
  \frac{1}{n} \sum_{k=1}^n x^2_k \leq \const{P}. \label{eq:powerdirect}
\end{equation}
Indeed, all achievability results (which are based on schemes that
ignore the feedback) are derived under
\eqref{eq:powerdirect}, whereas all converse results are derived under
\eqref{eq:power}. Since all mappings $\phi_n$ and
$\varphi_n^{(1)},\ldots,\varphi_n^{(n)}$ that satisfy
\eqref{eq:powerdirect} also fulfill \eqref{eq:power}, this implies that
the achievability results as well as the converse results derived in
this paper hold irrespective of whether constraint \eqref{eq:power} or
\eqref{eq:powerdirect} is imposed.
\end{remark}

\section{Channel Capacity}
\label{sec:capacity}
Let the \emph{rate} $R$ (in nats per channel use) be defined as
\begin{equation}
  R \triangleq \frac{\log |\set{M}|}{n},
\end{equation}
where $\log(\cdot)$ denotes the natural
logarithm function.
A rate is said to be \emph{achievable} if there exists a sequence
of mappings $\{\phi_n\}$ (without feedback) or $\bigl\{\bigl(\varphi_n^{(1)},\ldots,\varphi_n^{(n)}\bigr)\bigr\}$ (with feedback)
and $\{\psi_n\}$ such that the error probability $\Prob\big(\hat{M} \neq M\big)$
tends to zero as $n$ goes to infinity. The \emph{capacity} $C$ is the
supremum of all achievable rates. We denote by $C(\SNR)$ the capacity
under the input constraint \eqref{eq:power} when there is no feedback, and we add the subscript ``FB''
to indicate that there is a feedback link. Clearly
\begin{equation}
  C(\SNR) \leq C_{\textnormal{FB}}(\SNR) \label{eq:noFBtoFB}
\end{equation}
as we can always ignore the feedback link.

In the absence of feedback, the \emph{information capacity} is defined
as
\begin{equation}
  C_{\textnormal{Info}}(\SNR) \triangleq \varliminf_{n \to \infty}\frac{1}{n} \sup
  I(X_1^n;Y_1^n), \label{eq:info}
\end{equation}
where the supremum is over all joint distributions on $X_1,\ldots,X_n$ satisfying
\eqref{eq:power}. When there is a feedback link, then we define the
information capacity as
\begin{equation}
  C_{\textnormal{Info,FB}}(\SNR) \triangleq \varliminf_{n \to \infty}\frac{1}{n} \sup
  I(M;Y_1^n),
\end{equation}
where the supremum is over
all mappings $\varphi^{(1)}_n,\ldots,\varphi^{(n)}_n$ satisfying \eqref{eq:power}.
By Fano's inequality \cite[Thm.~2.11.1]{coverthomas91} no rate above
$C_{\textnormal{Info}}(\SNR)$ and $C_{\textnormal{Info,FB}}(\SNR)$ is
achievable, i.e.,
\begin{equation}
  C(\SNR) \leq
  C_{\textnormal{Info}}(\SNR) \qquad \textnormal{and} \qquad
  C_{\textnormal{FB}}(\SNR) \leq C_{\textnormal{Info,FB}}(\SNR). \label{eq:fano}
\end{equation}
See \cite{verduhan94} for conditions that guarantee that
$C_{\textnormal{Info}}(\SNR)$ is achievable. Note that the channel
\eqref{eq:channel} is not stationary\footnote{By a \emph{stationary channel} we mean a channel
  where for any stationary sequence of channel inputs $\{X_k\}$ and
  corresponding channel outputs $\{Y_k\}$ the pair $\{(X_k,Y_k)\}$ is
  jointly stationary.} since the variance of the additive noise
depends on the time-index $k$. It is therefore \emph{prima facie} not
clear whether the inequalities in \eqref{eq:fano} hold with equality.

In this paper, we shall investigate the capacities $C(\SNR)$ and
$\CFB$ at low SNR and at high SNR. To study capacity at low SNR, we
compute the \emph{capacities per unit cost} defined as \cite{verdu90}
\begin{equation}
  \dot{C}(0) \triangleq \sup_{\SNR>0}\frac{C(\SNR)}{\SNR} \qquad
  \textnormal{and} \qquad \Cfb \triangleq \sup_{\SNR>0} \frac{\CFB}{\SNR}.\label{eq:unitcostdef}
\end{equation}
It will become apparent later that the suprema in
\eqref{eq:unitcostdef} are attained when $\SNR$ tends to zero.
Note that \eqref{eq:noFBtoFB} implies
\begin{equation}
  \dot{C}(0) \leq \Cfb. \label{eq:FBtonoFB}
\end{equation}
At high SNR, we study
conditions under which capacity is unbounded in the SNR. Notice that when the
allowed transmit power is large, then there is a trade-off between
optimizing the present transmission and minimizing the interference to future
transmissions. Indeed, increasing the transmission power may help to
overcome the present ambient noise, but it also heats up the chip and
thus increases the noise variance in future receptions. \emph{Prima facie} it
is not clear that, as we increase the allowed transmit power, the
capacity tends to infinity. We shall see that this is not necessarily
the case.

\section{Main Results}
\label{sec:result}
Our main results are presented in the following two
sections. Section~\ref{sub:unitcost} focuses on capacity at low SNR
and presents our results on the capacity per unit cost. Section~\ref{sub:isbounded}
provides a sufficient condition and a necessary condition on
$\{\alpha_{\ell}\}$ under which capacity is bounded in the $\SNR$.

\subsection{Capacity per Unit Cost}
\label{sub:unitcost}
The results presented in this section hold under the additional assumptions that
\begin{equation}
  \label{eq:alpha}
  \sum_{\ell=1}^{\infty} \alpha_{\ell} \triangleq \alpha < \infty
\end{equation}
and that $\{U_k\}$ is IID.

\begin{proposition}
  \label{prop:lowSNR}
  Consider the above channel model, and assume additionally that
  the sequence $\{\alpha_{\ell}\}$ satisfies \eqref{eq:alpha} and that
  $\{U_k\}$ is IID. Then
  \begin{equation}
    \sup_{\SNR >0} \frac{C_{\textnormal{Info}}(\SNR)}{\SNR} \geq
    \sup_{\SNR>0} \frac{C_{\alpha=0}(\SNR)}{\SNR},
  \end{equation}
  where $C_{\alpha=0}(\SNR)$ denotes the capacity of the
  channel
  \begin{equation*}
    Y_k = x_k + \sigma\cdot U_k
  \end{equation*}
  which is a special case of \eqref{eq:channel} for $\alpha=0$.
\end{proposition}
\begin{proof}
  See Appendix~\ref{app:proplowSNR}.
\end{proof}
This proposition demonstrates that the heating up can only increase
the information capacity per unit cost. Thus at low SNR the
heating effect is unharmful.

For \emph{Gaussian} noise, i.e., if
$\{U_k\}$ is a sequence of IID, zero-mean, unit-variance
\emph{Gaussian} random variables, then the heating effect is
beneficial.
\begin{theorem}
  \label{thm:lowSNR}
  Consider the above channel model, and assume additionally that
  the sequence $\{\alpha_{\ell}\}$ satisfies \eqref{eq:alpha} and that
  $\{U_k\}$ is a sequence of IID, zero-mean, unit-variance Gaussian
  random variables. Then,
  irrespective of whether feedback is available or not, the
  corresponding capacity per unit cost is given by
  \begin{equation}
    \Cfb = \dot{C}(0) = \lim_{\SNR \downarrow 0} \frac{C(\SNR)}{\SNR}
    = \frac{1}{2}\left(1+\sum_{\ell=1}^{\infty}\alpha_{\ell}\right). \label{eq:thmlowSNR}
  \end{equation}
\end{theorem}
\begin{proof}
  See Section~\ref{sec:lowSNR}.
\end{proof}
For example, for the geometric heat dissipation profile
\eqref{eq:geometric} we obtain from Theorem~\ref{thm:lowSNR}
\begin{equation}
  \Cfb = \dot{C}(0) = \frac{1}{2}\frac{1}{1-\rho}, \qquad 0<\rho<1.
\end{equation}
Thus the capacity per unit cost is monotonically \emph{decreasing} in $\rho$.

The above result might be counterintuitive, because it suggests not to use
heat sinks at low SNR. Nevertheless it can be heuristically explained by
noting that the heating effect increases the \emph{channel
  gain}\footnote{The channel gain
is given by the ratio of the ``desired'' power at the channel output
to the ``desired'' power at the channel input.}. Indeed, if we split up
the channel output
\begin{equation*}
  Y_k = X_k+\sqrt{\left(\sigma^2+\sum_{\ell=1}^{k-1}\alpha_{k-\ell}X_{\ell}^2\right)}\cdot U_k
\end{equation*}
into a data-dependent part
\begin{IEEEeqnarray*}{lCl}
  \tilde{X}_k & = &
  X_k+\sqrt{\left(\sum_{\ell=1}^{k-1}\alpha_{k-\ell}X_{\ell}^2\right)}\cdot
  U_k
\end{IEEEeqnarray*}
and a data-independent part $Z_k$ (with $\{Z_k\}$ being a sequence of IID, zero-mean,
variance-$\sigma^2$, Gaussian random variables drawn independently of
$\{(U_k,X_k)\}$), then the channel gain $\const{G}$ for \eqref{eq:channel} is given by
\begin{equation}
  \const{G} \triangleq \lim_{n \to \infty} \sup \frac{\sum_{k=1}^n
  \E{\tilde{X}_k^2}}{\sum_{k=1}^n \E{X_k^2}} = 1+
  \sum_{\ell=1}^{\infty} \alpha_{\ell}, \label{eq:inview}
\end{equation}
where the supremum is over all joint distributions on $X_1,\ldots,X_n$
satisfying \eqref{eq:power}. Thus, in view of \eqref{eq:inview}, Theorem~\ref{thm:lowSNR} demonstrates that
the capacity per unit cost is determined by the channel gain
$\const{G}$.
This result is not specific to \eqref{eq:channel} but has also been observed
for other channel models. For example,
the same is true for fading channels whenever the
additive noise is Gaussian \cite{verdu02},
\cite{lapidothshamai02}.

\subsection{Conditions for Bounded Capacity}
\label{sub:isbounded}
While at low SNR the heating effect is
beneficial, at high SNR it is detrimental. In fact,
it turns out that capacity can be even bounded in
the SNR, i.e., the capacity does not tend to infinity as the SNR tends
to infinity. The following theorem provides a sufficient condition and a
necessary condition on $\{\alpha_{\ell}\}$ for the capacity to be
bounded. Note that the results presented in this section do not
require the additional assumptions made in
Section~\ref{sub:unitcost}: we neither assume that the
sequence $\{\alpha_{\ell}\}$ satisfies \eqref{eq:alpha} nor that
$\{U_k\}$ is IID.

\begin{theorem}
  Consider the channel model described in
  Section~\ref{sec:channelmodel}. Then
  \begin{IEEEeqnarray}{llCl}
    \textnormal{i)} \quad & \left(\varliminf_{\ell \to \infty}
    \frac{\alpha_{\ell+1}}{\alpha_{\ell}} > 0\right) \quad &\Longrightarrow &
    \quad  \left(\sup_{\SNR > 0} C_{\textnormal{FB}}(\SNR)
    < \infty\right)  \label{eq:main1}\\
    \textnormal{ii)} & \left(\varlimsup_{\ell \to \infty}
    \frac{\alpha_{\ell+1}}{\alpha_{\ell}} = 0\right)\quad & \Longrightarrow
    & \quad  \left(\sup_{\SNR > 0} C(\SNR) = \infty\right), \label{eq:main2}
  \end{IEEEeqnarray}
  where we define, for any $a>0$, $a/0 \triangleq \infty$ and $0/0\triangleq 0$.
  \label{thm:highSNR}
\end{theorem}
\begin{proof}
  See Section~\ref{sec:highSNR}.
\end{proof}
For example, for a geometric heat dissipation \eqref{eq:geometric} we
have
\begin{equation*}
  \lim_{\ell \to \infty} \frac{\alpha_{\ell+1}}{\alpha_{\ell}} = \rho
  , \qquad 0<\rho<1 
\end{equation*}
and it follows from Theorem~\ref{thm:highSNR} that the corresponding capacity is bounded. On the other hand,
for a sub-geometric heat dissipation, i.e.,
\begin{equation*}
  \alpha_{\ell} = \rho^{\ell^{\kappa}}, \qquad \ell \in \Integers^+
\end{equation*}
for some $0<\rho<1$ and $\kappa>1$, we obtain
\begin{equation*}
  \lim_{\ell \to \infty} \frac{\alpha_{\ell+1}}{\alpha_{\ell}} =
  \lim_{\ell \to \infty} \rho^{(\ell+1)^{\kappa}-\ell^{\kappa}} = 0
\end{equation*}
and Theorem~\ref{thm:highSNR} implies that the corresponding capacity
is unbounded. Roughly speaking, we can say that whenever the sequence of
coefficients $\{\alpha_{\ell}\}$ decays \emph{not faster than
geometrically} then capacity is \emph{bounded} in the SNR, and whenever
the sequence of coefficients $\{\alpha_{\ell}\}$ decays \emph{faster than
  geometrically} then capacity is \emph{unbounded} in the SNR.

\begin{remark}
For Part i) of Theorem~\ref{thm:highSNR} the assumptions that the process
$\{U_k\}$ is weakly-mixing and that it has a finite fourth moment are
not needed. These assumptions are only needed in the proof of Part
ii).\footnote{They are needed to prove
  Lemma~\ref{lemma:typical}.} In Part ii) of
Theorem~\ref{thm:highSNR}, the condition on the left-hand side (LHS)
of \eqref{eq:main2} can be replaced by
\begin{equation}
  \lim_{\ell \to \infty}
  \frac{1}{\ell}\log\frac{1}{\alpha_{\ell}} = \infty. \label{eq:newcond}
\end{equation}
This condition \eqref{eq:newcond} is weaker than the original
condition \eqref{eq:main2} because
\begin{equation*}
  \left(\varlimsup_{\ell\to\infty} \frac{\alpha_{\ell+1}}{\alpha_{\ell}} = 0\right)
  \quad \Longrightarrow \quad \left(\lim_{\ell\to\infty}
  \frac{1}{\ell}\log\frac{1}{\alpha_{\ell}} = \infty\right).
\end{equation*}
\end{remark}

When neither the LHS of \eqref{eq:main1} nor the LHS of
\eqref{eq:main2} hold, i.e.,
\begin{equation}
  \varlimsup_{\ell \to \infty} \frac{\alpha_{\ell+1}}{\alpha_{\ell}} >
  0 \qquad \textnormal{and} \qquad \varliminf_{\ell \to \infty}
  \frac{\alpha_{\ell+1}}{\alpha_{\ell}} = 0,  \label{eq:beyond}
\end{equation}
then capacity can be bounded or unbounded. Example~\ref{ex:1} exhibits
a sequence $\{\alpha_{\ell}\}$ satisfying \eqref{eq:beyond} for which
the capacity is bounded, and Example~\ref{ex:2} provides a sequence
$\{\alpha_{\ell}\}$ satisfying \eqref{eq:beyond} for which the
capacity is unbounded.\footnote{The provided sequences $\{\alpha_{\ell}\}$ are not monotonically
  decreasing in $\ell$. Consequently, Examples~\ref{ex:1} \& \ref{ex:2} are
  rather of mathematical than of practical interest. Nevertheless
  they show that when neither condition of
  Theorem~\ref{thm:highSNR} is satisfied, then one can construct
  simple examples yielding a bounded capacity or an unbounded
  capacity, thus demonstrating the difficulty of finding conditions
  that are necessary \emph{and} sufficient for the capacity to be
  bounded.}

\begin{example}
  \label{ex:1}
  Consider the sequence $\{\alpha_{\ell}\}$ where all
  coefficients with an even index are equal to $1$, and where all coefficients with an
  odd index are $0$. It satisfies \eqref{eq:beyond} because $\varlimsup_{\ell \to \infty}
  \alpha_{\ell+1}/\alpha_{\ell} = \infty$
  and $\varliminf_{\ell \to \infty} \alpha_{\ell+1}/\alpha_{\ell} =
  0$. Then the time-$k$ channel output $Y_k$ corresponding to the
  channel inputs $(x_1,\ldots,x_k)$ is given by
  \begin{equation*}
    Y_k = x_k + \sqrt{\left(\sigma^2+\sum_{\ell=1}^{\lfloor
    (k-1)/2\rfloor} x_{k-2\ell}^2\right)}\cdot U_k, \qquad k\in\Integers^+,
  \end{equation*}
  where $\lfloor \cdot \rfloor$ denotes the floor function.
  Thus at even times the output $Y_{2k}$, $k\in\Integers^+$ only depends on the ``even''
  inputs $(X_2,X_4,\ldots,X_{2k})$, while at odd times the output $Y_{2k+1}$, $k \in
  \Integers^+_0$ only depends on the ``odd'' inputs
  $(X_1,X_3,\ldots,X_{2k+1})$. By proceeding along the lines of the
  proof of Part i) of Theorem~\ref{thm:highSNR} while choosing in
  \eqref{eq:cauchy} $\beta=1/y_{k-2}^2$, it can be shown that the
  capacity of this channel is bounded.\footnote{Intuitively, with this
  choice of $\{\alpha_{\ell}\}$ the channel can be divided into two
  parallel channels, one connecting the inputs and outputs at even
  times, and the other connecting the inputs and outputs at odd times. As
  both channels have the coefficients
  $\tilde{\alpha}_0=\tilde{\alpha}_1=\ldots=1$, it follows from
  Theorem~\ref{thm:highSNR} that the capacity of each parallel channel
  is bounded and therefore also the capacity of the original channel.}
\end{example}

\begin{example}
  \label{ex:2}
  Consider the sequence $\{\alpha_{\ell}\}$ where all coefficients
  with an even positive index are $0$, and where all other
  coefficients are
  $1$. (Again, we have $\varlimsup_{\ell \to \infty}
  \alpha_{\ell+1}/\alpha_{\ell}=\infty$ and $\varliminf_{\ell \to
  \infty} \alpha_{\ell+1}/\alpha_{\ell}=0$.) In this case the
  time-$k$ channel output $Y_k$ corresponding to $(x_1,\ldots,x_k)$ is
  given by
  \begin{equation*}
    Y_k = x_k + \sqrt{\left(\sigma^2+\sum_{\ell=1}^{\lfloor
          k/2\rfloor} x_{k-2\ell+1}^2\right)}\cdot U_k, \qquad k\in\Integers^+.
  \end{equation*}
  Using Gaussian inputs of power
  $2\const{P}$ at even times while setting the inputs to be zero at
  odd times, and measuring the channel outputs only at even times,
  reduces the channel to a memoryless additive noise channel and
  demonstrates (using the result of \cite{lapidoth96}) the
  achievability of
  \begin{equation*}
     R = \frac{1}{4} \log(1+2\:\SNR)
  \end{equation*}
  which is unbounded in the $\SNR$.
\end{example}

The two seemingly-similar examples thus lead to completely different
capacity results. The crucial difference between Example~\ref{ex:1}
and Example~\ref{ex:2} is that in the former example at even times the
interference is caused by the past channel inputs at \emph{even} times,
whereas in the latter example at even times the interference 
is caused by the past channel inputs at \emph{odd} times. Thus in
Example~\ref{ex:2} setting all ``odd'' inputs to
zero cancels (at even times) the interference from past channel
inputs and hence transforms the channel into an
additive noise channel whose capacity is unbounded. Evidently, this
approach does not work for Example~\ref{ex:1}.

\section{Proof of Theorem~\ref{thm:lowSNR}}
\label{sec:lowSNR}

In Section~\ref{sub:upperLOW} we derive an upper bound on the feedback
capacity $C_{\textnormal{FB}}(\SNR)$, and in Section~\ref{sub:lowerLOW} we
derive a lower bound on the capacity $C(\SNR)$ in the absence of
feedback. These bounds are used in Section~\ref{sub:asymptotic} to derive an upper bound on $\Cfb$ and a lower
bound on $\dot{C}(0)$, which are then both shown to be
equal to $1/2\: (1+\alpha)$. Together with \eqref{eq:FBtonoFB} this
proves Theorem~\ref{thm:lowSNR}.

\subsection{Converse}
\label{sub:upperLOW}
The upper bound on $C_{\textnormal{FB}}(\SNR)$ is based on
\eqref{eq:fano} and on an upper bound on $\frac{1}{n} I(M;Y_1^n)$,
which for our channel can be expressed, using the chain rule for
mutual information, as
\begin{IEEEeqnarray}{lCl}
  \frac{1}{n} I(M;Y_1^n)
  & = & \frac{1}{n}\sum_{k=1}^n
  \Bigl(h\big(Y_k\big|Y_1^{k-1}\big)-h\big(Y_k\big|Y_1^{k-1},M\big)\Bigr)\nonumber\\
  & = & \frac{1}{n}\sum_{k=1}^n
  \Bigl(h\big(Y_k\big|Y_1^{k-1}\big)-h\big(Y_k\big|Y_1^{k-1},M,X_1^k\big)\Bigr)\nonumber\\
  & = & \frac{1}{n}\sum_{k=1}^n
  \Bigg(h\big(Y_k\big|Y_1^{k-1}\big)-h(U_k) -\frac{1}{2}\E{\log\left(\sigma^2+\sum_{\ell=1}^{k-1}\alpha_{k-\ell}X_{\ell}^2\right)}\Bigg),
  \IEEEeqnarraynumspace \label{eq:upper1}
\end{IEEEeqnarray}
where the second equality follows because $X_1^k$ is a function of
$M$ and $Y_1^{k-1}$; and the last equality follows from the behavior
of
differential entropy under translation and scaling \cite[Thms.~9.6.3
\& 9.6.4]{coverthomas91}, and because $U_k$
is independent of $\big(Y_1^{k-1},M,X_1^k\big)$.

Evaluating the differential entropy $h(U_k)$ of a Gaussian random
variable, and using the trivial lower bound $\E{\log\left(\sigma^2+\sum_{\ell=1}^{k-1}\alpha_{k-\ell}X_{\ell}^2\right)}
  \geq \log \sigma^2$,
we obtain the final upper bound
\begin{IEEEeqnarray}{lCl}
  \frac{1}{n} I(M;Y_1^n)
  & \leq & \frac{1}{n}\sum_{k=1}^n
  \left(h\big(Y_k\big|Y_1^{k-1}\big)-\frac{1}{2}\log(2\pi
      e\sigma^2)\right)\nonumber\\
  & \leq & \frac{1}{n}\sum_{k=1}^n \frac{1}{2}
  \log\left(1+\sum_{\ell=1}^{k}\alpha_{k-\ell} \E{X_{\ell}^2}/\sigma^2\right)\nonumber\\
  & \leq &
  \frac{1}{2}\log\left(1+\frac{1}{n}\sum_{k=1}^n\sum_{\ell=1}^k\alpha_{k-\ell}
  \E{X_{\ell}^2}/\sigma^2\right)\nonumber\\
  & = &
  \frac{1}{2}\log\left(1+\frac{1}{n}\sum_{k=1}^n \E{X_{k}^2}/\sigma^2\sum_{\ell=0}^{n-k}\alpha_{\ell}\right)\nonumber\\
  & \leq &
  \frac{1}{2}\log\left(1+(1+\alpha)\:\frac{1}{n}\sum_{k=1}^n \E{X_{k}^2}/\sigma^2\right)\nonumber\\
  & \leq & \frac{1}{2}\log\left(1+(1+\alpha)\:\SNR\right),\label{eq:upper2}
\end{IEEEeqnarray}
where we define $\alpha_0 \triangleq 1$. Here the second inequality
follows because conditioning cannot increase entropy and from the entropy maximizing property of Gaussian
random variables \cite[Thm.~9.6.5]{coverthomas91}; the next inequality follows by Jensen's inequality;
the following equality by rewriting the double sum; the subsequent inequality
follows because the coefficients are nonnegative which implies that
$\sum_{\ell=0}^{n-k}\alpha_{\ell} \leq \sum_{\ell=0}^{\infty}
\alpha_{\ell}=1+\alpha$; and the last inequality follows from the power
constraint \eqref{eq:power}.

\subsection{Direct Part}
\label{sub:lowerLOW}
As aforementioned, the above channel \eqref{eq:channel} is not
stationary and it is therefore \emph{prima facie} not clear
whether $C_{\textnormal{Info}}(\SNR)$ is achievable.
We shall sidestep this problem by studying the capacity of a different channel whose time-$k$
channel output $\tilde{Y}_k\in\Reals$ is, conditional on the sequence
$\{X_k\}=\{x_k\}$, given by
\begin{equation}
  \tilde{Y}_{k} = x_k +
  \sqrt{\left(\sigma^2+\sum_{\ell=-\infty}^{k-1}\alpha_{k-\ell}x_{\ell}^2\right)}\cdot
  U_k, \qquad k\in\Integers^+, \label{eq:newchannel}
\end{equation}
where $\{U_k\}$ and $\{\alpha_{\ell}\}$ are defined in
Section~\ref{sec:channelmodel}. This channel has the advantage that it
is stationary \& ergodic in the sense that when $\{X_k\}$ is a stationary \& ergodic process then the
pair $\{(X_k,\tilde{Y}_k)\}$ is jointly stationary \& ergodic.
It follows that if the sequences
$\{X_k\,,\,k=0,-1,\ldots\}$ and $\{X_k\, ,\, k=1,2,\ldots\}$ are independent of each
other, and if the random variables $X_k$, $k=0,-1,\ldots$ are bounded, then any rate
that can be achieved over this new channel is also achievable over the
original channel. Indeed, the original channel \eqref{eq:channel} can be
converted into
\eqref{eq:newchannel} by adding
\begin{equation*}
  S_k = \sqrt{\left(\sum_{\ell=-\infty}^0 \alpha_{k-\ell}
    X_{\ell}^2\right)}\cdot U_{-k}
\end{equation*}
to the channel output $Y_k$,\footnote{The
  boundedness of the random variables $X_k$, $k=0,-1,\ldots$ guarantees
  that the quantity $\sum_{\ell=-\infty}^0\alpha_{k-\ell}x_{\ell}^2$ is
  finite for any realization of $\{X_k\, , \, k =0,-1,\ldots\}$.} and,
  since the independence of $\{X_k\, ,\,k=0,-1,\ldots\}$ and
$\{X_k\, ,\,k=1,2,\ldots\}$ ensures that the sequence $\{S_k\, ,\, k\in\Integers^+\}$ is independent of
the message $M$, it follows that any rate achievable over
\eqref{eq:newchannel} can be achieved over \eqref{eq:channel} by using
a receiver that generates $\{S_k\, ,\, k\in\Integers^+\}$ and
guesses then $M$ based on $(Y_1+S_1,\ldots,Y_n+S_n)$.\footnote{Note that
  this approach is specific to the case where $\{U_k\}$ is a sequence of
  Gaussian random variables. Indeed, it relies heavily on the fact
  that given $\{X_k\}=\{x_k\}$ the additive noise term on the
  right-hand side of \eqref{eq:newchannel} can be
  written as the sum of two independent random variables, of which one
  only depends on $\{X_k\, ,\,k=0,-1,\ldots\}$ and the other only on
  $\{X_k\, ,\,k=1,2,\ldots\}$. This surely
  holds for Gaussian random variables, but it does not necessarily
  hold for other distributions on $\{U_k\}$.}

We shall consider channel inputs $\{X_{k}\}$ that are blockwise IID in blocks of $L$
symbols (for some $L\in\Integers^+$). Thus denoting
$\vect{X}_{b}=\trans{(X_{b L +1},\ldots,X_{(b+1)L})}$ (where
$\trans{(\cdot)}$ denotes the transpose), $\{\vect{X}_{b}\}$ is a
sequence of IID random length-$L$ vectors with
$\vect{X}_{b}$ taking on the value $\trans{(\xi,0,\ldots,0)}$ with
probability $\delta$ and $\trans{(0,\ldots,0)}$ with
probability $1-\delta$, for some $\xi \in \Reals$. Note that to
  satisfy the average-power constraint \eqref{eq:power} we shall
  choose $\xi$ and $\delta$ so that
\begin{equation}
  \frac{\xi^2}{\sigma^2} \delta = L \: \SNR. \label{eq:LBpower}
\end{equation}

Let $\tilde{\vect{Y}}_{b}=\trans{(\tilde{Y}_{b L
    +1},\ldots,\tilde{Y}_{(b+1)L})}$. Noting that the pair $\{(\vect{X}_{b},\tilde{\vect{Y}}_{b})\}$ is jointly
stationary \& ergodic, it follows from \cite{verduhan94} that the rate
\begin{equation*}
  \lim_{n \to \infty} \frac{1}{n}
  I\Bigl(\vect{X}_0^{\lfloor n/L \rfloor -1};\tilde{\vect{Y}}_0^{\lfloor
      n/L \rfloor-1}\Bigr)
\end{equation*}
is achievable over the new channel \eqref{eq:newchannel} and thus
yields a lower bound on the capacity $C(\SNR)$ of the original
channel \eqref{eq:channel}.
We lower bound
$\frac{1}{n}I\big(\vect{X}_0^{\lfloor n/L \rfloor -1};\tilde{\vect{Y}}_0^{\lfloor
    n/L \rfloor-1}\big)$ as
\begin{IEEEeqnarray}{lCl}
  \frac{1}{n} I\big(\vect{X}_0^{\lfloor n/L \rfloor -1};\tilde{\vect{Y}}_0^{\lfloor
    n/L \rfloor-1}\big)
  & = & \frac{1}{n} \sum_{b=0}^{\lfloor n/L \rfloor -1}
  I\big(\vect{X}_{b};\tilde{\vect{Y}}_0^{\lfloor n/L \rfloor
    -1}\big|\vect{X}_0^{b-1}\big)\nonumber\\
  & \geq & \frac{1}{n} \sum_{b=0}^{\lfloor n/L \rfloor -1}
  I\big(\vect{X}_{b};\tilde{\vect{Y}}_{b}\big|\vect{X}_0^{b-1}\big)\nonumber\\
  & \geq & \frac{1}{n} \sum_{b=0}^{\lfloor n/L \rfloor -1}\left(
  I\big(\vect{X}_{b};\tilde{\vect{Y}}_{b}\big|\vect{X}_{-\infty}^{b-1}\big)-I\big(\vect{X}_{-\infty}^{-1};\tilde{\vect{Y}}_{b}\big|\vect{X}_0^{b}\big)\right), \label{eq:LB1}
\end{IEEEeqnarray}
where we use the chain rule and the nonnegativity of mutual
information. It is shown in Appendix~\ref{app:tozero}
that
\begin{equation}
  \lim_{b \to \infty}
  I\big(\vect{X}_{-\infty}^{-1};\tilde{\vect{Y}}_{b}\big|\vect{X}_0^{b}\big)
  = 0.
\end{equation}
This together with a Ces\'aro type theorem
\cite[Thm.~4.2.3]{coverthomas91} yields
\begin{IEEEeqnarray}{lCl}
  \lim_{n \to \infty} \frac{1}{n} I\big(\vect{X}_0^{\lfloor n/L \rfloor -1};\tilde{\vect{Y}}_0^{\lfloor
    n/L \rfloor-1}\big)
  & \geq & 
    \frac{1}{L}
    I\big(\vect{X}_{0};\tilde{\vect{Y}}_{0}\big|\vect{X}_{-\infty}^{-1}\big)
    -\frac{1}{L}\lim_{n \to \infty}
  \frac{1}{\lfloor n/L \rfloor} \sum_{b=0}^{\lfloor n/L \rfloor -1}I\big(\vect{X}_{-\infty}^{-1};\tilde{\vect{Y}}_{b}\big|\vect{X}_0^{b}\big)
  \nonumber\\
  & = & \frac{1}{L} I\big(\vect{X}_{0};\tilde{\vect{Y}}_{0}\big|\vect{X}_{-\infty}^{-1}\big),\label{eq:LBcesaro}
\end{IEEEeqnarray}
where the first inequality follows by the stationarity of
$\{(\vect{X}_{b},\tilde{\vect{Y}}_{b})\}$ which implies that
$I\big(\vect{X}_{b};\tilde{\vect{Y}}_{b}|\vect{X}_{-\infty}^{b-1}\big)$
does not depend on $b$, and by noting that \mbox{$\lim_{n \to \infty}
\frac{\lfloor n/L \rfloor}{n}=1/L$}.

We proceed to analyze
$I\big(\vect{X}_{0};\tilde{\vect{Y}}_{0}|\vect{X}_{-\infty}^{-1}=\vect{x}_{-\infty}^{-1}\big)$
for a given sequence $\vect{X}_{-\infty}^{-1}=\vect{x}_{-\infty}^{-1}$. Making
use of the canonical decomposition of mutual information (e.g., \cite[Eq.~(10)]{verdu90}), we have
\begin{IEEEeqnarray}{lCl}
  I\big(\vect{X}_{0};\tilde{\vect{Y}}_{0}\big|\vect{X}_{-\infty}^{-1}=\vect{x}_{-\infty}^{-1}\big)
  & = & I\big(X_{1};\tilde{\vect{Y}}_{0}\big|\vect{X}_{-\infty}^{-1}=\vect{x}_{-\infty}^{-1}\big)
  \nonumber\\
  & = & \int D\Big( P_{\tilde{\vect{Y}}_{0}|X_{1}=x,\vect{x}_{-\infty}^{-1}} \Big\| P_{\tilde{\vect{Y}}_{0}|X_{1}=0,\vect{x}_{-\infty}^{-1}} \Big) \d  P_{X_{1}}(x)\nonumber\\
  & & {} -D\Big(P_{\tilde{\vect{Y}}_{0}|\vect{x}_{-\infty}^{-1}} \Big\|
    P_{\tilde{\vect{Y}}_{0}|X_{1}=0,\vect{x}_{-\infty}^{-1}}\Big)\nonumber\\
  & = & \delta D\Big(P_{\tilde{\vect{Y}}_{0}|X_{1}=\xi,\vect{x}_{-\infty}^{-1}} \Big\| P_{\tilde{\vect{Y}}_{0}|X_{1}=0,\vect{x}_{-\infty}^{-1}}\Big) \nonumber\\
  & & {} - D\Big(P_{\tilde{\vect{Y}}_{0}|\vect{x}_{-\infty}^{-1}} \Big\|
    P_{\tilde{\vect{Y}}_{0}|X_{1}=0,\vect{x}_{-\infty}^{-1}}\Big),
  \label{eq:LB2}
\end{IEEEeqnarray}
where the first equality
follows because, for our choice of input distribution, $X_{2}=\ldots=X_{L}=0$ and hence $X_{1}$ conveys as
much information about $\tilde{\vect{Y}}_{0}$ as $\vect{X}_{0}$.
Here $D(\cdot\|\cdot)$ denotes relative entropy, i.e.,
\begin{equation*}
  D\big(P_1\big\|P_0\big) = \left\{ \begin{array}{ll} \displaystyle
  \int \log \frac{\d P_1}{\d P_0} \d P_1 \quad & \displaystyle
  \textnormal{if } P_1
  \ll P_0 \\[8pt] \displaystyle +\infty & \displaystyle
  \textnormal{otherwise,} \end{array} \right.
\end{equation*}
 and
\begin{equation*}
P_{\tilde{\vect{Y}}_{0}|X_{1}=\xi,\vect{x}_{-\infty}^{-1}}, \quad
P_{\tilde{\vect{Y}}_{0}|X_{1}=0,\vect{x}_{-\infty}^{-1}},\quad
\textnormal{and}\quad 
P_{\tilde{\vect{Y}}_{0}|\vect{x}_{-\infty}^{-1}} 
\end{equation*}
denote the distributions
of $\tilde{\vect{Y}}_{0}$ conditional on the inputs
$\big(X_{1}=\xi,\vect{X}_{-\infty}^{-1}=\vect{x}_{-\infty}^{-1}\big)$,
$\big(X_{1}=0,\vect{X}_{-\infty}^{-1}=\vect{x}_{-\infty}^{-1}\big)$,
and on $\vect{X}_{-\infty}^{-1}=\vect{x}_{-\infty}^{-1}$, respectively. Thus
$P_{\tilde{\vect{Y}}_{0}|X_{1}=\xi,\vect{x}_{-\infty}^{-1}}$ is the
law of an
$L$-variate Gaussian random vector of mean $\trans{(\xi,0,\ldots,0)}$
and of diagonal covariance matrix $\mat{K}^{(\xi)}_{\vect{x}_{-\infty}^{-1}}$ with diagonal
entries
\begin{IEEEeqnarray*}{lCl}
  \mat{K}^{(\xi)}_{\vect{x}_{-\infty}^{-1}}(1,1) & = &
  \sigma^2+\sum_{\ell=-\infty}^{-1}\alpha_{-\ell L}x_{\ell L+1}^2 \\
  \mat{K}^{(\xi)}_{\vect{x}_{-\infty}^{-1}}(i,i) & = &
  \sigma^2+\alpha_{i-1}\xi^2+\sum_{\ell=-\infty}^{-1}\alpha_{-\ell
  L+i-1}x_{\ell L+1}^2,\qquad i=2,\ldots,L;
\end{IEEEeqnarray*}
$P_{\tilde{\vect{Y}}_{0}|X_{1}=0,\vect{x}_{-\infty}^{-1}}$ is
the law of an $L$-variate, zero-mean Gaussian random vector of diagonal
covariance matrix $\mat{K}^{(0)}_{\vect{x}_{-\infty}^{-1}}$ with diagonal entries
\begin{equation*}
  \mat{K}^{(0)}_{\vect{x}_{-\infty}^{-1}}(i,i) =
  \sigma^2+\sum_{\ell=-\infty}^{-1}\alpha_{-\ell L+i-1}x_{\ell L+1}^2, \quad i=1,\ldots,L;
\end{equation*}
and $P_{\tilde{\vect{Y}}_{0}|\vect{x}_{-\infty}^{-1}}$ is given by
\begin{equation*}
  P_{\tilde{\vect{Y}}_{0}|\vect{x}_{-\infty}^{-1}} = \delta P_{\tilde{\vect{Y}}_{0}|X_{1}=\xi,\vect{x}_{-\infty}^{-1}}+(1-\delta)P_{\tilde{\vect{Y}}_{0}|X_{1}=0,\vect{x}_{-\infty}^{-1}}.
\end{equation*}

In order to evaluate the first term on the right-hand side (RHS) of
\eqref{eq:LB2} we note that the relative entropy of two real, $L$-variate Gaussian
random vectors of means $\bfmu_1$ and $\bfmu_2$ and
of covariance matrices $\mat{K}_1$ and $\mat{K}_2$ is
given by
\begin{IEEEeqnarray}{lCl}
  D\bigl(\Normal{\bfmu_1}{\mat{K}_1}\bigm\|\Normal{\bfmu_2}{\mat{K}_2}\bigr)
  & = & \frac{1}{2} \log\det\mat{K}_2 -
  \frac{1}{2}\log\det\mat{K}_1 +
  \frac{1}{2}\trace{\mat{K}_1\mat{K}_2^{-1}-\mat{I}_L} \nonumber\\
  & & {} +
  \frac{1}{2}\trans{(\bfmu_1-\bfmu_2)}\mat{K}_2^{-1}(\bfmu_1-\bfmu_2), \label{eq:DGaussian}
\end{IEEEeqnarray}
with $\det\mat{A}$ and $\trace{\mat{A}}$ denoting
the determinant and the trace of the matrix $\mat{A}$,
and where $\mat{I}_L$ denotes the $L \times L$ identity matrix. The
second term on the RHS of \eqref{eq:LB2} is analyzed in the next
subsection. 

Let $\E{D\big(P_{\tilde{\vect{Y}}_{0}|\vect{X}_{-\infty}^{-1}}
    \big\|P_{\tilde{\vect{Y}}_{0}|X_{1}=0,\vect{X}_{-\infty}^{-1}}\big)}$ denote the second term on
the RHS of \eqref{eq:LB2} averaged over $\vect{X}_{-\infty}^{-1}$,
i.e.,
\begin{IEEEeqnarray}{c}
  \E{D\Big(P_{\tilde{\vect{Y}}_{0}|\vect{X}_{-\infty}^{-1}}
        \Big\|P_{\tilde{\vect{Y}}_{0}|X_{1}=0,\vect{X}_{-\infty}^{-1}}\Big)}
        = \E[\vect{X}_{-\infty}^{-1}]{D\Big(P_{\tilde{\vect{Y}}_{0}|\vect{x}_{-\infty}^{-1}}
    \Big\|P_{\tilde{\vect{Y}}_{0}|X_{1}=0,\vect{x}_{-\infty}^{-1}}\Big)}.\nonumber
\end{IEEEeqnarray}
Then using \eqref{eq:DGaussian} \& \eqref{eq:LB2} and taking expectations
over $\vect{X}_{-\infty}^{-1}$, we obtain, again defining $\alpha_0
\triangleq 1$,
\begin{IEEEeqnarray}{lCl}
  \frac{1}{L}
    I\big(\vect{X}_{0};\tilde{\vect{Y}}_{0}\big|\vect{X}_{-\infty}^{-1}\big)
  & = & \frac{\delta}{L}\frac{\xi^2}{\sigma^2}\frac{1}{2}\sum_{i=1}^L
  \E{\frac{\alpha_{i-1}}{1+\sum_{\ell=-\infty}^{-1}
      \alpha_{-\ell L+i-1} X_{\ell L+1}^2/\sigma^2}}\nonumber\\
  & & {} -\frac{\delta}{L}\frac{1}{2}\sum_{i=2}^L
  \E{\log\left(1+\frac{\alpha_{i-1}\xi^2}{\sigma^2+\sum_{\ell=-\infty}^{-1}\alpha_{-\ell L+i-1}X_{\ell L+1}^2}\right)}
  \nonumber\\
  & & {} - \frac{1}{L}\E{D\Big(P_{\tilde{\vect{Y}}_{0}|\vect{X}_{-\infty}^{-1}} \Big\|
      P_{\tilde{\vect{Y}}_{0}|X_{1}=0,\vect{X}_{-\infty}^{-1}}\Big)}\nonumber\\
  & \geq & \frac{\delta}{L}\frac{\xi^2}{\sigma^2}\frac{1}{2}\sum_{i=1}^L
  \frac{\alpha_{i-1}}{1+\sum_{\ell=-\infty}^{-1}
      \alpha_{-\ell L+i-1} \E{X_{\ell L+1}^2}/\sigma^2}\nonumber\\
  & & {} -\frac{\delta}{L}\frac{1}{2}\sum_{i=2}^L
  \log\left(1+\alpha_{i-1} \xi^2/\sigma^2\right)
  \nonumber\\
  & & {} - \frac{1}{L}\E{D\Big(P_{\tilde{\vect{Y}}_{0}|\vect{X}_{-\infty}^{-1}} \Big\|
      P_{\tilde{\vect{Y}}_{0}|X_{1}=0,\vect{X}_{-\infty}^{-1}}\Big)}\nonumber\\
  & \geq & \frac{1}{2} \SNR \sum_{i=1}^L
  \frac{\alpha_{i-1}}{1+ \alpha \: L\:\SNR}\nonumber\\
  & & {} -\frac{1}{2}\SNR \sum_{i=2}^L
  \frac{\log\left(1+\alpha_{i-1} \xi^2/\sigma^2\right)}{\xi^2/\sigma^2}
  \nonumber\\
  & & {} - \frac{1}{L}\E{D\Big(P_{\tilde{\vect{Y}}_{0}|\vect{X}_{-\infty}^{-1}} \Big\|
      P_{\tilde{\vect{Y}}_{0}|X_{1}=0,\vect{X}_{-\infty}^{-1}}\Big)}, \label{eq:LBbeforelimit}
\end{IEEEeqnarray}
where the first inequality follows by the lower bound $\E{1/(1+X)}
\geq 1/(1+\E{X})$, which is a consequence of Jensen's inequality applied to
the convex function $1/(1+x)$, $x>0$, and by the upper bound
\begin{IEEEeqnarray*}{c}
  \E{\log\left(1+\frac{\alpha_{i-1}\xi^2}{\sigma^2+\sum_{\ell=-\infty}^{-1}\alpha_{-\ell
  L+i-1}X_{\ell L+1}^2}\right)} \leq \log\left(1+\alpha_{i-1}
  \xi^2/\sigma^2\right), \qquad i=2,\ldots,L;
\end{IEEEeqnarray*}
and the second inequality
follows by \eqref{eq:LBpower} and by upper bounding
\begin{equation*}
  \sum_{\ell=-\infty}^{-1} \alpha_{-\ell L+i-1} \leq \sum_{\ell=1}^{\infty}
  \alpha_{\ell} = \alpha, \qquad i=1,\ldots,L.
\end{equation*}

The final lower bound follows now by \eqref{eq:LBbeforelimit} and \eqref{eq:LBcesaro}
\begin{IEEEeqnarray}{lCl}
  \lim_{n \to
  \infty}\frac{1}{n}I\big(\vect{X}_0^{\lfloor n/L
  \rfloor-1};\tilde{\vect{Y}}_0^{\lfloor n/L \rfloor -1}\big)
  & \geq & \frac{1}{2} \SNR \sum_{i=1}^L
  \frac{\alpha_{i-1}}{1+ \alpha \: L \:\SNR} \nonumber\\
  & & {} - \frac{1}{2}\SNR \sum_{i=2}^L
    \frac{\log\left(1+\alpha_{i-1}  \xi^2/\sigma^2\right)}{\xi^2/\sigma^2}
    \nonumber\\
    & & - \frac{1}{L} \E{D\Big(P_{\tilde{\vect{Y}}_{0}|\vect{X}_{-\infty}^{-1}} \Big\|
        P_{\tilde{\vect{Y}}_{0}|X_{1}=0,\vect{X}_{-\infty}^{-1}}\Big)} \label{eq:LBfinal}
\end{IEEEeqnarray}
and by recalling that
\begin{equation}
  C(\SNR) \geq \lim_{n \to
  \infty}\frac{1}{n}I\big(\vect{X}_0^{\lfloor n/L
  \rfloor-1};\tilde{\vect{Y}}_0^{\lfloor n/L \rfloor -1}\big).
\end{equation}

\subsection{Asymptotic Analysis}
\label{sub:asymptotic}
We start with analyzing the upper bound \eqref{eq:upper2}. Using that $\log(1+x) \leq
x$, $x>-1$ we have
\begin{equation}
  \frac{C_{\textnormal{FB}}(\SNR)}{\SNR} \leq
  \frac{\frac{1}{2}\log(1+(1+\alpha)\: \SNR)}{\SNR}\leq \frac{1}{2}(1+\alpha),
\end{equation}
and we thus obtain
\begin{equation}
  \Cfb = \sup_{\SNR>0} \frac{C_{\textnormal{FB}}(\SNR)}{\SNR} \leq \frac{1}{2}(1+\alpha).\label{eq:asymU}
\end{equation}

In order to derive a lower bound on $\dot{C}(0)$ we first note that
\begin{equation}
  \dot{C}(0) = \sup_{\SNR >0}\frac{C(\SNR)}{\SNR} \geq \lim_{\SNR
  \downarrow 0} \frac{C(\SNR)}{\SNR} \label{eq:concave}
\end{equation}
and proceed by analyzing the limiting ratio of the lower bound
\eqref{eq:LBfinal} to SNR as SNR tends to zero.
To this end we first shall show that
\begin{equation}
  \lim_{\SNR \downarrow 0} \frac{\E{D\Big(P_{\tilde{\vect{Y}}_{0}|\vect{X}_{-\infty}^{-1}} \Big\|
        P_{\tilde{\vect{Y}}_{0}|X_{1}=0,\vect{X}_{-\infty}^{-1}}\Big)}}{\SNR} = 0. \label{eq:o(snr)}
\end{equation}
We recall that for any pair of distributions $P_0$
and $P_1$ satisfying $P_1 \ll P_0$ \cite[p.~1023]{verdu90}
\begin{equation}
  \lim_{\beta \downarrow 0} \frac{D\left(\left.\beta P_1 + (1-\beta) P_0\right\|
      P_0\right)}{\beta} =0.\label{eq:anydensities}
\end{equation}
Thus, for any given $\vect{X}_{-\infty}^{-1}=\vect{x}_{-\infty}^{-1}$,
\eqref{eq:anydensities} together with $\delta = \SNR \:L\:
\sigma^2/\xi^2$ implies that
\begin{equation}
  \lim_{\SNR \downarrow 0} \frac{D\Big(P_{\tilde{\vect{Y}}_{0}|\vect{x}_{-\infty}^{-1}} \Big\|
      P_{\tilde{\vect{Y}}_{0}|X_{1}=0,\vect{x}_{-\infty}^{-1}}\Big)}{\SNR} = 0. \label{eq:forany}
\end{equation}
In order to show that this also holds when $D\Big(P_{\tilde{\vect{Y}}_{0}|\vect{x}_{-\infty}^{-1}} \Big\|
  P_{\tilde{\vect{Y}}_{0}|X_{1}=0,\vect{x}_{-\infty}^{-1}}\Big)$ is averaged over
$\vect{X}_{-\infty}^{-1}$,
we derive in the following the uniform upper bound
\begin{IEEEeqnarray}{c}
  \sup_{\vect{x}_{-\infty}^{-1}} D\Big(P_{\tilde{\vect{Y}}_{0}|\vect{x}_{-\infty}^{-1}} \Big\|
      P_{\tilde{\vect{Y}}_{0}|X_{1}=0,\vect{x}_{-\infty}^{-1}}\Big)
    = \left.D\Big(P_{\tilde{\vect{Y}}_{0}|\vect{x}_{-\infty}^{-1}} \Big\|
      P_{\tilde{\vect{Y}}_{0}|X_{1}=0,\vect{x}_{-\infty}^{-1}}\Big)\right|_{\vect{x}_{-\infty}^{-1}=0}. \IEEEeqnarraynumspace \label{eq:uniformbound}
\end{IEEEeqnarray}
The claim \eqref{eq:o(snr)} follows
then by upper bounding
\begin{IEEEeqnarray}{c}
  \E{D\Big(P_{\tilde{\vect{Y}}_{0}|\vect{X}_{-\infty}^{-1}} \Big\|
        P_{\tilde{\vect{Y}}_{0}|X_{1}=0,\vect{X}_{-\infty}^{-1}}\Big)} \leq \left.D\Big(P_{\tilde{\vect{Y}}_{0}|\vect{x}_{-\infty}^{-1}} \Big\|
      P_{\tilde{\vect{Y}}_{0}|X_{1}=0,\vect{x}_{-\infty}^{-1}}\Big)\right|_{\vect{x}_{-\infty}^{-1}=0} \IEEEeqnarraynumspace
\end{IEEEeqnarray}
and by \eqref{eq:forany}.

In order to prove \eqref{eq:uniformbound} we use that any Gaussian random
vector can be expressed as the sum of two independent Gaussian random vectors to
write the channel output $\tilde{\vect{Y}}_{0}$ as
\begin{equation}
  \tilde{\vect{Y}}_{0} = \vect{X}_{0}+\vect{V}+\vect{W},
\end{equation}
where, conditional on $\vect{X}_{-\infty}^{0}=\vect{x}_{-\infty}^{0}$, $\vect{V}$
and $\vect{W}$ are $L$-variate,
zero-mean Gaussian random vectors, drawn independently of each other
and having
the respective diagonal covariance matrices
$\mat{K}_{\vect{V}|\vect{x}_{0}}$ and
$\mat{K}_{\vect{W}|\vect{x}_{-\infty}^{-1}}$
whose diagonal entries are given by
\begin{IEEEeqnarray*}{lCl}
  \mat{K}_{\vect{V}|\vect{x}_{0}}(1,1) & = & \sigma^2\\
  \mat{K}_{\vect{V}|\vect{x}_{0}}(i,i) & = & \sigma^2 +
  \alpha_{i-1}x_{1}^2, \qquad i=2,\ldots,L,
\end{IEEEeqnarray*}
and
\begin{equation*}
  \mat{K}_{\vect{W}|\vect{x}_{-\infty}^{-1}}(i,i) = \sum_{\ell=-\infty}^{-1}
  \alpha_{-\ell L+i-1}x_{\ell L+1}^2, \qquad i=1,\ldots,L.
\end{equation*}
Thus $\vect{W}$ is the portion of the noise due
to $\vect{X}_{-\infty}^{-1}$, and $\vect{V}$ is the portion of the
noise that remains after subtracting $\vect{W}$.
Note that $\vect{X}_{0}+\vect{V}$ and $\vect{W}$ are
independent of each other because $\vect{X}_{0}$ is, by construction, independent
of $\vect{X}_{-\infty}^{-1}$.
The upper bound \eqref{eq:uniformbound} follows now by
\begin{IEEEeqnarray}{lCl}
  D\Big(P_{\tilde{\vect{Y}}_{0}|\vect{x}_{-\infty}^{-1}} \Big\|
      P_{\tilde{\vect{Y}}_{0}|X_{1}=0,\vect{x}_{-\infty}^{-1}}\Big)
  & = & D\Big(P_{\vect{X}_{0}+\vect{V}+\vect{W}|\vect{x}_{-\infty}^{-1}} \Big\|
    P_{\vect{X}_{0}+\vect{V}+\vect{W}|X_{1}=0,\vect{x}_{-\infty}^{-1}}\Big)\nonumber\\
  & \leq & D\big(P_{\vect{X}_{0}+\vect{V}} \big\|
    P_{\vect{X}_{0}+\vect{V}|X_{1}=0}\big)\nonumber\\
  & = & \left.D\Big(P_{\tilde{\vect{Y}}_{0}|\vect{x}_{-\infty}^{-1}} \Big\|
      P_{\tilde{\vect{Y}}_{0}|X_{1}=0,\vect{x}_{-\infty}^{-1}}\Big)\right|_{\vect{x}_{-\infty}^{-1}=0},\label{eq:asym1}
\end{IEEEeqnarray}
where
\begin{equation*}
P_{\vect{X}_{0}+\vect{V}+\vect{W}|\vect{x}_{-\infty}^{-1}} \quad \textnormal{and}
\quad P_{\vect{X}_{0}+\vect{V}+\vect{W}|X_{1}=0,\vect{x}_{-\infty}^{-1}}
\end{equation*} 
denote the distributions of $\vect{X}_{0}+\vect{V}+\vect{W}$
conditional on the inputs $\vect{X}_{-\infty}^{-1}=\vect{x}_{-\infty}^{-1}$
and on $(X_{1}=0,\vect{X}_{-\infty}^{-1}=\vect{x}_{-\infty}^{-1})$,
respectively; $P_{\vect{X}_{0}+\vect{V}}$ denotes
the unconditional distribution of
$\vect{X}_{0}+\vect{V}$; and $P_{\vect{X}_{0}+\vect{V}|X_{1}=0}$ denotes the distribution of
$\vect{X}_{0}+\vect{V}$ conditional on $X_{1}=0$.
Here the
inequality follows by the data processing inequality for relative
entropy (see \cite[Sec.~2.9]{coverthomas91}) and by noting that
$\vect{X}_{0}+\vect{V}$ is independent of $\vect{X}_{-\infty}^{-1}$.

Returning to the analysis of \eqref{eq:LBfinal}, we obtain from
\eqref{eq:concave} and \eqref{eq:o(snr)}
\begin{IEEEeqnarray}{lCl}
  \dot{C}(0) & \geq & \lim_{\SNR\downarrow 0} \frac{C(\SNR)}{\SNR}\nonumber\\
  & \geq & \lim_{\SNR \downarrow 0} \frac{1}{2} \sum_{i=1}^L
  \frac{\alpha_{i-1}}{1+ \alpha \: L \:\SNR} - \frac{1}{2} \sum_{i=2}^L
  \frac{\log\left(1+\alpha_{i-1}\xi^2/\sigma^2\right)}{\xi^2/\sigma^2}\nonumber\\
  & = & \frac{1}{2} \sum_{i=1}^{L} \alpha_{i-1} - \frac{1}{2} \sum_{i=2}^L
  \frac{\log\left(1+\alpha_{i-1}
  \xi^2/\sigma^2\right)}{\xi^2/\sigma^2}.
\end{IEEEeqnarray}
By letting first $\xi^2$ go to infinity while holding $L$ fixed, and by letting
then $L$ go to infinity, we obtain the desired lower bound on the
capacity per unit cost
\begin{equation}
  \dot{C}(0) \geq \lim_{\SNR\downarrow 0}\frac{C(\SNR)}{\SNR}\geq \frac{1}{2} (1+\alpha).\label{eq:asymL}
\end{equation}
Thus \eqref{eq:asymL}, \eqref{eq:FBtonoFB}, and \eqref{eq:asymU} yield
\begin{equation}
  \frac{1}{2} (1+\alpha) \leq \lim_{\SNR\downarrow 0}\frac{C(\SNR)}{\SNR}\leq \dot{C}(0) \leq \Cfb \leq \frac{1}{2}(1+\alpha)
\end{equation}
which proves Theorem~\ref{thm:lowSNR}.

\section{Proof of Theorem~\ref{thm:highSNR}}
\label{sec:highSNR}

\subsection{Part i)}
\label{sub:bounded}
In order to show that
\begin{equation}\varliminf_{\ell \to
  \infty} \frac{\alpha_{\ell+1}}{\alpha_{\ell}}>0 \label{eq:condi}
\end{equation}
implies that the feedback capacity
$\CFB$ is bounded, we derive a capacity upper bound which is based on
\eqref{eq:fano} and on an upper bound on
$\frac{1}{n}I(M;Y_1^n)$. Again we define $\alpha_0\triangleq 1$.

We first note that, according to
\eqref{eq:condi}, we can find an $\ell_0 \in \Integers^+$ and a $0<\rho<1$ so
that 
\begin{equation}
  \alpha_{\ell_0} > 0 \qquad \textnormal{and} \qquad
  \frac{\alpha_{\ell+1}}{\alpha_{\ell}} \geq \rho, \quad \ell\geq\ell_0. \label{eq:rho}
\end{equation}
We continue with the chain rule for mutual information
\begin{IEEEeqnarray}{lCl}
  \frac{1}{n} I(M;Y_1^n)
  & = & \frac{1}{n} \sum_{k=1}^{\ell_0}
  I\big(M;Y_k\big|Y_1^{k-1}\big)+\frac{1}{n}\sum_{k=\ell_0+1}^n
  I\big(M;Y_k\big|Y_1^{k-1}\big).\IEEEeqnarraynumspace \label{eq:firstsum}
\end{IEEEeqnarray}
Each summand in the first sum on the RHS of
\eqref{eq:firstsum} is upper bounded by
\begin{IEEEeqnarray}{lCl}
  I\big(M;Y_k\big|Y_1^{k-1}\big)
  & \leq & h(Y_k) - h\big(Y_k\big|Y_1^{k-1},M\big) \nonumber\\
  & = & h(Y_k)-\frac{1}{2}\E{\log\left(\sigma^2+\sum_{\ell=1}^{k-1}
      \alpha_{k-\ell}X_{\ell}^2\right)} - h\big(U_k\big|U_1^{k-1}\big)\nonumber\\
  & \leq & \frac{1}{2}\log\left(2\pi e\left(1+\sum_{\ell=1}^k \alpha_{k-\ell}
    \frac{\E{X_{\ell}^2}}{\sigma^2}\right)\!\right) - h\big(U_k\big|U_1^{k-1}\big)
  \nonumber\\
  & \leq & \frac{1}{2}\log\left(2\pi e\left(1+ \big(\sup_{\ell' \in \Integers^+_0}
  \alpha_{\ell'}\big) \sum_{\ell=1}^k
    \frac{\E{X_{\ell}^2}}{\sigma^2}\right)\!\right) - h\big(U_k\big|U_1^{k-1}\big)
  \nonumber\\
  & \leq & \frac{1}{2}\log\left(2\pi e\left(1+\big(\sup_{\ell' \in \Integers^+_0}
      \alpha_{\ell'}\big) \:n\:
      \SNR\right)\!\right)-h\big(U_k\big|U_1^{k-1}\big) \nonumber\\
  & \leq & \frac{1}{2}\log\left(2\pi e\left(1+\big(\sup_{\ell' \in \Integers^+_0}
      \alpha_{\ell'}\big) \:n\:
      \SNR\right)\!\right)-h\big(U_k\big|U_{-\infty}^{k-1}\big).
\IEEEeqnarraynumspace\label{eq:firstterm}
\end{IEEEeqnarray}
Recall that $\sup_{\ell' \in \Integers^+_0}
\alpha_{\ell'}$ is finite \eqref{eq:bounded}. Here the first inequality follows because conditioning cannot
increase entropy; the following equality follows because $\big(X_1^k,U_1^{k-1}\big)$ is
a function of $\big(M,Y_1^{k-1}\big)$, from the behavior of entropy under
translation and scaling \cite[Thms.~9.6.3 \& 9.6.4]{coverthomas91},
and from the fact that, conditional on $U_1^{k-1}$, $U_k$ is
independent of $\big(X_1^k,M,Y_1^{k-1}\big)$; the subsequent inequality follows
from the entropy maximizing property of Gaussian random variables
 and by lower bounding \mbox{$\E{\log\left(\sigma^2+\sum_{\ell=1}^{k-1}
      \alpha_{k-\ell}X_{\ell}^2\right)} \geq \log\sigma^2$};
the next inequality by upper bounding each coefficient
\mbox{$\alpha_{\ell} \leq \sup_{\ell' \in \Integers_0^+} \alpha_{\ell'}$}, $\ell=1,\ldots,k$;
the subsequent inequality follows from the power constraint \eqref{eq:power}; and the last
inequality follows because conditioning cannot increase entropy.

The summands in the second sum on the RHS of
\eqref{eq:firstsum} are upper bounded using the general upper bound
for mutual information \cite[Thm.~5.1]{lapidothmoser03_3}
\begin{equation}
  I(X;Y) \leq \int D\big(W(\cdot|x) \big\| R(\cdot)\big) \d Q(x), \label{eq:duality}
\end{equation}
where $W(\cdot|\cdot)$ is the channel law, $Q(\cdot)$ is the
distribution on the channel input $X$, and $R(\cdot)$ is any
distribution on the output alphabet. Thus any choice of output
distribution $R(\cdot)$ yields an upper bound on the mutual
information.

We upper bound $I\big(M;Y_k\big|Y_1^{k-1}=y_1^{k-1}\big)$, $k=\ell_0+1,\ldots,n$
for a given $Y_1^{k-1}=y_1^{k-1}$
by choosing $R(\cdot)$ to be a Cauchy distribution whose density
is given by
\begin{equation}
  \frac{\sqrt{\beta}}{\pi}\frac{1}{1+\beta y_k^2},
  \qquad y_k \in \Reals, \label{eq:cauchy}
\end{equation}
where we choose the scale parameter $\beta$ to be\footnote{When $y_{k-\ell_0}=0$
  then with this choice of $\beta$ the density of the Cauchy distribution \eqref{eq:cauchy} is
  undefined. However, this event is of zero probability and has
  therefore no impact on the mutual information $I\big(M;Y_k\big|Y_1^{k-1}\big)$.}
\begin{equation}
\beta = \frac{1}{\tilde{\beta} y_{k-\ell_0}^2} \qquad \textnormal{and} \qquad
\tilde{\beta} =
  \min\left\{\rho^{\ell_0-1}\:\frac{\alpha_{\ell_0}}{\displaystyle
  \max_{\ell'=0,\ldots,\ell_0-1} \alpha_{\ell'}},\alpha_{\ell_0},\rho^{\ell_0}\right\}, \label{eq:beta}
\end{equation}
with $0<\rho<1$ and $\ell_0\in\Integers^+$ given by \eqref{eq:rho}. 
Note that \eqref{eq:rho} together with \eqref{eq:bounded} implies that
\begin{equation}
  0<\tilde{\beta}<1 \qquad \textnormal{and} \qquad \tilde{\beta} \alpha_{\ell} \leq \alpha_{\ell+\ell_0}, \quad
  \ell \in \Integers^+_0. \label{eq:betaalpha}
\end{equation}
Applying \eqref{eq:cauchy} to \eqref{eq:duality} yields
\begin{IEEEeqnarray}{lCl}
  I\big(M;Y_k\big|Y_1^{k-1}=y_1^{k-1}\big)
  & \leq &
  \Econd{\log\left(1+\frac{Y_k^2}{\tilde{\beta}Y_{k-\ell_0}^2}\right)}{Y_1^{k-1}=y_1^{k-1}}+\frac{1}{2}
  \log \big(\tilde{\beta}y_{k-\ell_0}^2\big)\nonumber\\
  & & {} + \log
  \pi - h\big(Y_k\big|M,Y_1^{k-1}=y_1^{k-1}\big),
\end{IEEEeqnarray}
and we thus obtain, averaging over $Y_1^{k-1}$,
\begin{IEEEeqnarray}{lCl}
  I\big(M;Y_k\big|Y_1^{k-1}\big)
  & \leq & \log\pi -
  h\big(Y_k\big|Y_1^{k-1},M\big) + \frac{1}{2}
  \E{\log\big(\tilde{\beta}Y_{k-\ell_0}^2\big)}\nonumber\\
  & & {} + \E{\log\big(\tilde{\beta}Y_{k-\ell_0}^2+Y_k^2\big)}-\E{\log\big(Y^2_{k-\ell_0}\big)} -\log\tilde{\beta}.\IEEEeqnarraynumspace \label{eq:UB1}
\end{IEEEeqnarray}

We evaluate the terms on the RHS of \eqref{eq:UB1} individually. We
begin with
\begin{equation}
  h\big(Y_k\big|Y_1^{k-1},M\big) \geq \frac{1}{2}\E{\log\left(\sigma^2+\sum_{\ell=1}^{k-1}
      \alpha_{k-\ell} X_{\ell}^2\right)} + h\big(U_k\big|U_{-\infty}^{k-1}\big),
  \label{eq:1}
\end{equation}
where we use the same steps as in the equality in \eqref{eq:firstterm}
and that conditioning cannot increase entropy. The next term is
upper bounded by
\begin{IEEEeqnarray}{lCl}
    \E{\log\big(\tilde{\beta}Y_{k-\ell_0}^2\big)}
  & = &
    \E{\Econd{\log\left(\tilde{\beta}\big(X_{k-\ell_0}+\theta\big(X_1^{k-\ell_0-1}\big)\cdot
    U_{k-\ell_0}\big)^2\right)}{X_1^{k-\ell_0}}}
    \nonumber\\
    & \leq &
    \E{\log\left(\tilde{\beta}\Econd{\big(X_{k-\ell_0}+\theta\big(X_1^{k-\ell_0-1}\big)\cdot
    U_{k-\ell_0}\big)^2}{X_1^{k-\ell_0}}\right)}
  \nonumber\\
  & = & \E{\log\left(\tilde{\beta}X_{k-\ell_0}^2 +
    \tilde{\beta}\sigma^2+\tilde{\beta}\sum_{\ell=1}^{k-\ell_0-1}\alpha_{k-\ell_0-\ell}
    X_{\ell}^2\right)} \nonumber\\
& \leq & \E{\log\left(\sigma^2+\sum_{\ell=1}^{k-\ell_0}\alpha_{k-\ell}X_{\ell}^2\right)},\label{eq:2} 
\end{IEEEeqnarray}
where we define, for a given $X_1^{k-1}=x_1^{k-1}$,
\begin{equation} 
  \theta\big(x_1^{k-1}\big) \triangleq \sqrt{\sigma^2+\sum_{\ell=1}^{k-1}\alpha_{k-\ell}x_{\ell}^2}.\label{eq:sigma}
\end{equation}
Here the first inequality in \eqref{eq:2} follows from Jensen's
inequality, and the second inequality follows from
\eqref{eq:betaalpha}.
Similarly we use Jensen's inequality along with \eqref{eq:betaalpha} to upper bound
\begin{IEEEeqnarray}{lCl}
  \E{\log\big(\tilde{\beta}Y_{k-\ell_0}^2+Y_k^2\big)}
  & \leq &
  \E{\log\left(\sigma^2+\sum_{\ell=1}^{k-\ell_0}\alpha_{k-\ell}X_{\ell}^2+\sigma^2+\sum_{\ell=1}^k
      \alpha_{k-\ell}X_{\ell}^2\right)} \nonumber\\
  & \leq & \log 2 + \E{\log\left(\sigma^2+\sum_{\ell=1}^k \alpha_{k-\ell}X_{\ell}^2\right)}.\label{eq:3}
\end{IEEEeqnarray}
In order to lower bound $\E{\log\big(Y_{k-\ell_0}^2\big)}$ we
need the following lemma:
\begin{lemma}
  \label{lemma:expectedlog}
  Let $X$ be a random variable of density
  $f_{X}(x)$, $x \in \Reals$. Then, for any $0<\delta\leq 1$ and $0 <
  \eta < 1$ we have
  \begin{equation}
    \sup_{c \in \Reals}
    \E{\log |X+c|^{-1} \cdot \I{|X+c| \leq \delta}} \leq \eps(\delta,\eta)
    +\frac{1}{\eta} h^-(X)
  \end{equation}
  where $\I{\cdot}$ denotes the indicator function\footnote{The
  indicator function $\I{\textnormal{statement}}$ takes on the value
  $1$ if the statement is true and $0$ otherwise.}; $h^-(X)$ is defined as
  \begin{equation}
    h^-(X) \triangleq \int_{\{x\in\Reals: f_X(x)>1\}} f_X(x)\log
    f_X(x) \d x;
  \end{equation}
  and where $\eps(\delta,\eta)>0$ tends to zero as $\delta \downarrow 0$.
\end{lemma}
\begin{proof}
  See \cite[Lemma 6.7]{lapidothmoser03_3}.
\end{proof}
We write the expectation as
\begin{IEEEeqnarray*}{lCl}
  \E{\log\big(Y_{k-\ell_0}^2\big)}
  & = &
  \E{\Econd{\log\left(X_{k-\ell_0}+\theta\big(X_1^{k-\ell_0-1}\big)\cdot
  U_{k-\ell_0}\right)^2}{X_1^{k-\ell_0}}}\IEEEeqnarraynumspace
\end{IEEEeqnarray*}
and lower bound the conditional expectation for a given $X_1^{k-\ell_0}=x_1^{k-\ell_0}$ by
\begin{IEEEeqnarray}{lCl}
  \IEEEeqnarraymulticol{3}{l}{\Econd{\log\left(X_{k-\ell_0}+\theta\big(X_1^{k-\ell_0-1}\big)\cdot
  U_{k-\ell_0}\right)^2}{X_1^{k-\ell_0}=x_1^{k-\ell_0}}}\nonumber\\
  \quad & = & \log \theta^2\big(x_1^{k-\ell_0-1}\big) - 2\:
  \Econd{\log\left|\frac{X_{k-\ell_0}}{\theta\big(X_1^{k-\ell_0-1}\big)}+U_{k-\ell_0}\right|^{-1}}{X_1^{k-\ell_0}=x_1^{k-\ell_0}}
  \nonumber\\
  & \geq & \log\theta^2\big(x_1^{k-\ell_0-1}\big) -2\eps(\delta,\eta) -\frac{2}{\eta}
  h^-(U_{k-\ell_0})+
  \log\delta^2  \label{eq:bla1}
\end{IEEEeqnarray}
for some $0<\delta\leq 1$ and $0<\eta<1$. Here the
inequality follows by splitting the conditional
expectation into the two expectations 
\begin{IEEEeqnarray}{lCl}
  \IEEEeqnarraymulticol{3}{l}{\Econd{\log\left|\frac{X_{k-\ell_0}}{\theta\big(X_1^{k-\ell_0-1}\big)}+U_{k-\ell_0}\right|^{-1}}{X_1^{k-\ell_0}=x_1^{k-\ell_0}}}\nonumber\\
   \qquad  & = &
   \Econd{\log\left|\frac{X_{k-\ell_0}}{\theta\big(X_1^{k-\ell_0-1}\big)}+U_{k-\ell_0}\right|^{-1}\cdot
     \I{\left|\frac{X_{k-\ell_0}}{\theta\big(X_1^{k-\ell_0-1}\big)}+U_{k-\ell_0}\right|\leq
       \delta}}{X_1^{k-\ell_0}=x_1^{k-\ell_0}}\nonumber\\ 
   & & {} + \Econd{\log\left|\frac{X_{k-\ell_0}}{\theta\big(X_1^{k-\ell_0-1}\big)}+U_{k-\ell_0}\right|^{-1}\cdot
     \I{\left|\frac{X_{k-\ell_0}}{\theta\big(X_1^{k-\ell_0-1}\big)}+U_{k-\ell_0}\right| >
       \delta}}{X_1^{k-\ell_0}=x_1^{k-\ell_0}}  \nonumber
\end{IEEEeqnarray}
and by
upper bounding then the first term 
on the RHS
using Lemma~\ref{lemma:expectedlog} and the second
term by $-\log\delta$.
Averaging \eqref{eq:bla1} over $X_1^{k-\ell_0}$ yields
\begin{IEEEeqnarray}{lCl}
  \E{\log\big(Y_{k-\ell_0}^2\big)}
  & \geq  & \E{\log\left(\sigma^2 + \sum_{\ell=1}^{k-\ell_0-1}
      \alpha_{k-\ell_0-\ell}X_{\ell}^2\right)}
  - 2\eps(\delta,\eta)-\frac{2}{\eta}h^-(U_{k-\ell_0})+\log\delta^2.\IEEEeqnarraynumspace \label{eq:4}
\end{IEEEeqnarray}
Note that, since $U_{k-\ell_0}$ is of unit variance, \eqref{eq:finiteentropy}
together with \cite[Lemma 6.4]{lapidothmoser03_3} implies that
$h^-(U_{k-\ell_0})$ is finite.

Turning back to the upper bound \eqref{eq:UB1} we obtain from \eqref{eq:1}, \eqref{eq:2},
\eqref{eq:3}, and \eqref{eq:4}
\begin{IEEEeqnarray}{lCl}
  \IEEEeqnarraymulticol{3}{l}{I\big(M;Y_k\big|Y_1^{k-1}\big)}\nonumber\\
  \quad & \leq & \log\pi  -  \frac{1}{2}\E{\log\left(\sigma^2+\sum_{\ell=1}^{k-1}
      \alpha_{k-\ell} X_{\ell}^2\right)} - h\big(U_k\big|U_{-\infty}^{k-1}\big) \nonumber\\
  & & {}
  +\frac{1}{2}\E{\log\left(\sigma^2+\sum_{\ell=1}^{k-\ell_0}\alpha_{k-\ell}X_{\ell}^2\right)}
  + \log 2 + \E{\log\left(\sigma^2+\sum_{\ell=1}^k
      \alpha_{k-\ell}X_{\ell}^2\right)} \nonumber\\
  & & {} -\E{\log\left(\sigma^2 + \sum_{\ell=1}^{k-\ell_0-1}
      \alpha_{k-\ell_0-\ell}X_{\ell}^2\right)}+ 2\eps(\delta,\eta)+ \frac{2}{\eta}
  h^-(U_{k-\ell_0})  -\log\delta^2 - \log\tilde{\beta}
  \;\nonumber\\
  & \leq & \E{\log\left(\sigma^2+\sum_{\ell=1}^k
      \alpha_{k-\ell}X_{\ell}^2\right)} - \E{\log\left(\sigma^2 + \sum_{\ell=1}^{k-\ell_0-1}
      \alpha_{k-\ell_0-\ell}X_{\ell}^2\right)} + \const{K}, \IEEEeqnarraynumspace
\label{eq:UB2}
\end{IEEEeqnarray}
where
\begin{equation}
  \const{K} \triangleq \log\frac{2\pi}{\tilde{\beta}\delta^2} -h\big(U_k\big|U_{-\infty}^{k-1}\big) + \frac{2}{\eta}h^-(U_{k-\ell_0})+
  2\eps(\delta,\eta)
\end{equation}
is a finite constant, and where the last inequality in \eqref{eq:UB2}
follows because for any $X_{k-\ell_0+1}^{k-1}=x_{k-\ell_0+1}^{k-1}$ we have $\sum_{\ell=1}^{k-\ell_0} \alpha_{k-\ell}
\:x_{\ell}^2 \leq \sum_{\ell=1}^{k-1} \alpha_{k-\ell}\:
x_{\ell}^2$. Note that $\const{K}$ does not depend on $k$ as
the process $\{U_k\}$ is stationary.

Turning back to the evaluation of the second sum on the RHS of \eqref{eq:firstsum}, we
use that for any sequences $\{a_k\}$ and $\{b_k\}$
\begin{IEEEeqnarray}{lCl}
  \sum_{k=\ell_0+1}^n (a_{k}-b_{k})
& = & \sum_{k=n-2\ell_0+1}^n (a_k-b_{k-n+3\ell_0}) +
 \sum_{k=\ell_0+1}^{n-2\ell_0}(a_{k}-b_{k+2\ell_0}).\IEEEeqnarraynumspace
 \label{eq:sums}
\end{IEEEeqnarray}
Defining
\begin{equation}
a_k \triangleq \E{\log\left(\sigma^2+\sum_{\ell=1}^k
      \alpha_{k-\ell}X_{\ell}^2\right)}
\end{equation}
and
\begin{equation}
  b_k \triangleq  \E{\log\left(\sigma^2 + \sum_{\ell=1}^{k-\ell_0-1}
      \alpha_{k-\ell_0-\ell}X_{\ell}^2\right)}
\end{equation}
we have for the first sum on the RHS of \eqref{eq:sums}
\begin{IEEEeqnarray}{lCl}
  \sum_{k=n-2\ell_0+1}^n
    (a_k-b_{k-n+3\ell_0}) 
  & = &
  \sum_{k=n-2\ell_0+1}^n \E{\log\left(\frac{\sigma^2+\sum_{\ell=1}^{k}
          \alpha_{k-\ell}X_{\ell}^2}{\sigma^2+\sum_{\ell=1}^{k-n+2\ell_0-1}
          \alpha_{k-n+2\ell_0-\ell}X_{\ell}^2}\right)}\nonumber\\
  & \leq & 2\ell_0 \log\left(1+\big(\sup_{\ell
        \in \Integers^+_0}\alpha_{\ell}\big) \: n \: \SNR \right)\IEEEeqnarraynumspace\label{eq:wholesum1}
\end{IEEEeqnarray}
which follows by lower bounding the denominator by $\sigma^2$, and by using then
Jensen's inequality together with the third and fourth inequality in
\eqref{eq:firstterm}.
For the second sum on the RHS of \eqref{eq:sums} we have
\begin{IEEEeqnarray}{lCl}
  \sum_{k=\ell_0+1}^{n-2\ell_0}(a_{k}-b_{k+2\ell_0})
  & = & \sum_{k=\ell_0+1}^{n-2\ell_0} \E{\log\left(\frac{\sigma^2+\sum_{\ell=1}^k
      \alpha_{k-\ell}X_{\ell}^2}{\sigma^2 + \sum_{\ell=1}^{k+\ell_0-1}
        \alpha_{k+\ell_0-\ell}X_{\ell}^2}\right)} \nonumber\\
  & \leq & \sum_{k=\ell_0+1}^{n-2\ell_0} \E{\log\left(\frac{\sigma^2+\sum_{\ell=1}^k
      \alpha_{k+\ell_0-\ell}X_{\ell}^2}{\sigma^2 + \sum_{\ell=1}^{k+\ell_0-1}
        \alpha_{k+\ell_0-\ell}X_{\ell}^2}\right)} - (n-3\ell_0) \log
  \tilde{\beta}\nonumber\\
  & \leq & -(n-3\ell_0)\log\tilde{\beta}, \label{eq:wholesum2}
\end{IEEEeqnarray}
where the first inequality follows by adding $\log\tilde{\beta}$ to
the expectation and by upper bounding then $\tilde{\beta} \alpha_{\ell} <
\alpha_{\ell+\ell_0}$, $\ell \in \Integers_0^+$ \eqref{eq:betaalpha}; and the
last inequality follows because for any given
$X_{k+1}^{k+\ell_0-1}=x_{k+1}^{k+\ell_0-1}$ we have
$\sum_{\ell=1}^{k} \alpha_{k+\ell_0-\ell} \:x_{\ell}^2 \leq
\sum_{\ell=1}^{k+\ell_0-1} \alpha_{k+\ell_0-\ell} \:x_{\ell}^2$.

We apply now \eqref{eq:UB2}, \eqref{eq:sums}, \eqref{eq:wholesum1},
and \eqref{eq:wholesum2} to upper bound
\begin{IEEEeqnarray}{lCl}
  \frac{1}{n}\sum_{\ell=\ell_0+1}^n
    I\big(M;Y_k\big|Y_1^{k-1}\big)
  & \leq & \frac{n-\ell_0}{n} \const{K} + \frac{2\ell_0}{n} \log\left(1+\big(\sup_{\ell
      \in \Integers^+_0}\alpha_{\ell}\big) \: n \: \SNR\right) -
    \frac{n-3\ell_0}{n} \log\tilde{\beta} \IEEEeqnarraynumspace \label{eq:UB3}
\end{IEEEeqnarray}
which together with \eqref{eq:firstsum} and
\eqref{eq:firstterm} yields
\begin{IEEEeqnarray}{lCl}
  \frac{1}{n} I(M;Y_1^n)
  & \leq & \frac{n-\ell_0}{n}
  \const{K}-\frac{n-3\ell_0}{n}\log\tilde{\beta}
  +\frac{\ell_0}{2n}\log (2\pi e) -
  \frac{\ell_0}{n} h\big(U_k\big|U_{-\infty}^{k-1}\big)\nonumber\\
  & & {} + \frac{\ell_0}{n}\frac{5}{2} \log\left(1+\big(\sup_{\ell
      \in \Integers^+_0}\alpha_{\ell}\big) \: n \: \SNR\right). \IEEEeqnarraynumspace
\end{IEEEeqnarray}
This converges to
$\const{K}-\log\tilde{\beta} < \infty$ as we let $n$ tend to infinity,
thus proving that
$\varliminf_{\ell \to \infty} \alpha_{\ell+1}/\alpha_{\ell} > 0$
implies that the capacity $\CFB$ is bounded in the $\SNR$.

\subsection{Part ii)}
\label{sub:unbounded}
We shall show that 
\begin{equation}
\lim_{\ell \to
  \infty}\frac{1}{\ell}\log\frac{1}{\alpha_{\ell}}=\infty \label{eq:weaker}
\end{equation}
implies that
the capacity $C(\SNR)$ in the absence of feedback is unbounded in
the SNR. Part ii) of Theorem~\ref{thm:highSNR} follows then by noting that
\begin{equation}
  \varlimsup_{\ell \to \infty} \frac{\alpha_{\ell+1}}{\alpha_{\ell}}
  = 0 \quad \Longrightarrow \quad \lim_{\ell \to \infty}
  \frac{1}{\ell} \log\frac{1}{\alpha_{\ell}} = \infty.
\end{equation}

We prove the claim by proposing a coding scheme that achieves
an unbounded rate. We first note that \eqref{eq:weaker}
implies that for any $0<\varrho<1$ we can find an $\ell_0 \in
\Integers^+$ so that
\begin{equation}
  \alpha_{\ell} < \varrho^{\ell}, \quad \ell=\ell_0,\ell\geq\ell_0. \label{eq:positivecoefficients}
\end{equation}

If there exists an $\ell_0 \in
\Integers^+$ so that $\alpha_{\ell}=0$, $\ell\geq\ell_0$, then
we can achieve the (unbounded) rate
\begin{equation}
  R = \frac{1}{2L} \log(1+L\: \SNR), \qquad L \geq \ell_0
\end{equation}
by a coding scheme where the channel inputs $\{X_{kL+1}\, ,\, k\in\Integers^+_0\}$ are IID,
zero-mean Gaussian random variables of variance $L\const{P}$, and where the other inputs are
deterministically zero. Indeed, by waiting $L$ time-steps, the chip's
temperature cools down to the ambient one so that the
noise variance is independent of the previous channel inputs and we
can achieve---after appropriate normalization---the capacity of the
additive white Gaussian noise (AWGN) channel \cite{lapidoth96}.

For the more general case \eqref{eq:positivecoefficients} we propose
the following encoding and decoding scheme.
Let $x_1^n(m)$, $m \in \set{M}$ denote the codeword sent out by the transmitter
that corresponds to the message $M=m$. We choose some $L \geq \ell_0$ and
generate the components
$x_{kL+1}(m)$, $m \in \set{M}$, $k = 0,\ldots,\lfloor n/L
  \rfloor-1$ independently of each other according
to a zero-mean Gaussian law of variance $\const{P}$. The other components are set to zero.\footnote{It follows from
  the weak law of large numbers that, for any $ m\in \set{M}$, $\frac{1}{n}\sum_{k=1}^n
  x_k^2(m)$ converges to $\const{P}/L$ in probability as $n$
  tends to infinity. This guarantees that the probability that a codeword does not
  satisfy the per-message power constraint
  \eqref{eq:powerdirect}---and hence also the average-power constraint
  \eqref{eq:power}---vanishes as $n$ tends
  to infinity.}
 
The receiver uses a \emph{nearest neighbor decoder} in order to guess
 $M$ based on the received sequence of channel outputs $y_1^n$. Thus it
 computes $ \|\vect{y}-\vect{x}(m')\|^2$ for each $m'\in\set{M}$
 and decides on the message that satisfies
 \begin{equation}
   \hat{M} = \arg \min_{m' \in \set{M}} \|\vect{y}-\vect{x}(m')\|^2,
 \end{equation}
where ties are resolved with a fair coin flip. Here, $\|\cdot\|$
denotes the Euclidean norm, and $\vect{y}$ and
$\vect{x}(m')$ denote the respective vectors
$\trans{(y_{1},y_{L+1},\ldots,y_{(\lfloor n/L \rfloor-1)L+1})}$ and
$\trans{(x_{1}(m'),x_{L+1}(m'),\ldots,x_{(\lfloor n/L \rfloor-1)L+1}(m'))}$.

We are interested in the average probability of error $\Prob\big(\hat{M}
  \neq M\big)$, averaged over
all codewords in the codebook, and averaged over all codebooks. By
the symmetry of the codebook construction, the probability of error
corresponding to the $m$-th message $\Prob \big(\hat{M} \neq M \,\big|\, M=m\big)$
does not depend on $m$, and we thus conclude that $\Prob \big(\hat{M}
  \neq M\big) = \Prob\big(\hat{M} \neq M\,\big|\, M=1\big)$. We further note that
\begin{equation}
  \Prob\big(\hat{M} \neq M\,\big|\, M=1\big) \leq \Prob\Biggl(\bigcup_{m'=2}^{|\set{M}|}
  \|\vect{Y}-\vect{X}(m')\|^2 \leq \|\vect{Z}\|^2\Biggm| M=1\Biggr), \label{eq:errorprob}
\end{equation}
where 
\begin{equation*}
\vect{Z} = \trans{\left(\theta\big(X_1(1)\big)\cdot
U_{1},\theta\big(X_1^{L}(1)\big)\cdot
U_{L+1},\ldots,\theta\big(X_1^{(\lfloor n/L \rfloor-1)L+1}(1)\big)\cdot
U_{(\lfloor n/L \rfloor-1)L+1}\right)}
\end{equation*}
which is,
conditional on $M=1$, equal to $\|\vect{Y}-\vect{X}(1)\|^2$.
In order to analyze \eqref{eq:errorprob} we need the following lemma.

\begin{lemma}
  \label{lemma:typical}
  Consider the channel described in Section~\ref{sec:channelmodel},
  and assume that $\{\alpha_{\ell}\}$ satisfies
  \eqref{eq:weaker}. Further assume that $\{X_{kL+1}\,,\,k\in\Integers_0^+\}$
  is a sequence of IID, zero-mean Gaussian random variables of variance
  $\const{P}$, and that $X_k=0$ if $k\mod L \neq 1$ (where $k\mod L$
  stands for the remainder upon diving $k$ by $L$). Let the set $\set{D}_{\eps}$ be defined as
  \begin{IEEEeqnarray}{lCll}
    \set{D}_{\eps} & \triangleq & \Bigg\{(\vect{y},\vect{z})\in
  \Reals^{\lfloor n/L\rfloor}\times \Reals^{\lfloor n/L\rfloor}:
  & \quad \left|\frac{1}{\lfloor n/L \rfloor} \|\vect{y}\|^2-(\sigma^2+\const{P}+\alpha^{(L)}\: \const{P}) \right| < \eps, \nonumber\\
  & & & \quad \left|\frac{1}{\lfloor n/L \rfloor}
    \|\vect{z}\|^2-(\sigma^2+\alpha^{(L)} \: \const{P}) \right| <
    \eps \qquad \quad \Bigg\}, \IEEEeqnarraynumspace \label{eq:setdef}
  \end{IEEEeqnarray}
  with $\alpha^{(L)}$ being defined as 
  \begin{equation}
    \alpha^{(L)} \triangleq \sum_{\ell=1}^{\infty} \alpha_{\ell L}. \label{eq:lemmaalphaL}
  \end{equation}
  Then
  \begin{equation}
    \lim_{n \to \infty} \Prob\big((\vect{Y},\vect{Z}) \in \set{D}_{\eps}\big) = 1
  \end{equation}
  for any $\eps > 0$.
\end{lemma}
\begin{proof}
  See Appendix~\ref{app:lemmatypical}.
\end{proof}

In order to upper bound the RHS of \eqref{eq:errorprob} we proceed
along the lines of \cite{lapidoth96}, \cite{lapidothshamai02}. We have
\begin{IEEEeqnarray}{lCl}
  \IEEEeqnarraymulticol{3}{l}{\Prob\Biggl(\bigcup_{m'=2}^{|\set{M}|}
  \|\vect{Y}-\vect{X}(m')\|^2 \leq \|\vect{Z}\|^2 \Biggm| M=1\Biggr)}\nonumber\\
\qquad & \leq & \Prob\big((\vect{Y},\vect{Z}) \notin \set{D}_{\eps}\big)+  \int_{\set{D}_{\eps}} \Prob
  \Biggl(\bigcup_{m'=2}^{|\set{M}|}
  \|\vect{y}-\vect{X}(m')\|^2 \leq
  \|\vect{z}\|^2 \,\Biggm| \,(\vect{y},\vect{z}), M=1 \Biggr) \d
  P(\vect{y},\vect{z}), \IEEEeqnarraynumspace \label{eq:typical1}
\end{IEEEeqnarray}
where we use that, by the symmetry of the codebook construction, the
law of $(\vect{Y},\vect{Z})$ does not depend on $M$.
It follows from Lemma~\ref{lemma:typical} that the first term on
the RHS of \eqref{eq:typical1} vanishes as $n$ tends to infinity.
Since the
codewords are independent of each other, conditional on $M=1$, the distribution of
$\vect{X}(m')$, $m'=2,\ldots,|\set{M}|$ does not depend on $(\vect{y},\vect{z})$. We upper bound the
second term on the RHS of \eqref{eq:typical1} by analyzing $\Prob\big(\|\vect{y}-\vect{X}(m')\|^2 \leq
  \|\vect{z}\|^2\,\big|\,(\vect{y},\vect{z}),M=1\big)$, $m'=2,\ldots,|\set{M}|$ and
by applying then the union of events bound.

For $m'=2,\ldots,|\set{M}|$, we have
\begin{IEEEeqnarray}{lCl}
  \IEEEeqnarraymulticol{3}{l}{\Prob\big(\|\vect{y}-\vect{X}(m')\|^2 \leq
      \|\vect{z}\|^2\,\big|\,(\vect{y},\vect{z})\big)}\nonumber\\
    \quad & \leq & \exp\Bigg\{-s \lfloor n/L
    \rfloor(\sigma^2+\alpha^{(L)}\: \const{P}+\eps) + \frac{s
      \|\vect{y}\|^2}{1-2s\const{P}} - \frac{1}{2} \lfloor n/L \rfloor
    \log(1-2s\const{P})\Bigg\}, \,\,\, (\vect{y},\vect{z})\in\set{D}_{\eps} \IEEEeqnarraynumspace  \label{eq:typical2}
\end{IEEEeqnarray}
for any $s<0$. This follows by upper bounding $\|\vect{z}\|^2$ by $\lfloor
n/L\rfloor(\sigma^2+\alpha^{(L)}\: \const{P}+\eps)$ and from Chernoff's bound
\cite[Sec.~5.4]{gallager68}. Using that, for $(\vect{y},\vect{z})\in\set{D}_{\eps}$, 
\begin{equation*}
\|\vect{y}\|^2 > \lfloor
n/L \rfloor(\sigma^2+\const{P}+\alpha^{(L)}\: \const{P}-\eps)
\end{equation*}
 it follows from the union of events bound and from \eqref{eq:typical2}
that \eqref{eq:typical1} goes to zero as $n$
tends to infinity if for some $s<0$ the rate $R$ satisfies
\begin{IEEEeqnarray}{lCl}
  R & < & \frac{s}{L}(\sigma^2+\alpha^{(L)}\: \const{P} +\eps) +
  \frac{1}{2L}\log(1-2s\const{P}) - \frac{s}{L}
  \frac{\sigma^2+\const{P}+\alpha^{(L)} \:
  \const{P}-\eps}{1-2s\const{P}}. \IEEEeqnarraynumspace
\end{IEEEeqnarray}
Thus choosing $s = - 1/2\cdot 1/(1+\alpha^{(L)}\: \const{P})$
yields that any rate below
\begin{IEEEeqnarray}{l}
  -\frac{1}{2L}\frac{\sigma^2+\alpha^{(L)}\:\const{P}+\eps}{1+\alpha^{(L)}\: \const{P}} + \frac{1}{2L}
  \log\left(1+\frac{\const{P}}{1+\alpha^{(L)}\:
      \const{P}}\right)\nonumber\\
  {} +\frac{1}{2L}\frac{\sigma^2+\const{P}+\alpha^{(L)}\:
  \const{P}-\eps}{1+\alpha^{(L)}\:
  \const{P}}\frac{1}{1+\frac{\const{P}}{1+\alpha^{(L)}\:
    \const{P}}} \label{eq:ach1}
\end{IEEEeqnarray}
is achievable. As $\const{P}$ tends to infinity this
converges to
\begin{equation}
  \frac{1}{2L}\log\left(1+\frac{1}{\alpha^{(L)}}\right) >
  \frac{1}{2L}\log\frac{1}{\alpha^{(L)}}. \label{eq:achievable}
\end{equation}

It remains to show that given \eqref{eq:positivecoefficients} we can
make $-\frac{1}{L}\log\alpha^{(L)}$ arbitrarily large. Indeed,
\eqref{eq:positivecoefficients} implies that
\begin{equation*}
  \alpha^{(L)} = \sum_{\ell=1}^{\infty} \alpha_{\ell L} <
  \sum_{\ell=1}^{\infty} \varrho^{\ell L} = \frac{\varrho^L}{1-\varrho^L} \label{eq:alphaL}
\end{equation*}
and \eqref{eq:achievable} can therefore be further lower bounded by
\begin{equation}
  \frac{1}{2L}\log\left(1-\varrho^L\right) + \frac{1}{2}\log\frac{1}{\varrho}.
\end{equation}
Letting $L$ tend to infinity yields then that we can achieve any rate
below $\frac{1}{2}\log\frac{1}{\varrho}$. As this can be made arbitrarily large by
choosing $\varrho$ sufficiently small, we conclude that
$\lim_{\ell \to \infty}
\frac{1}{\ell} \log\frac{1}{\alpha_{\ell}}=\infty$ implies that the
capacity is unbounded.

\section{Conclusion}
\label{sec:summary}
We studied a model for on-chip communication with nonideal heat
sinks. To account for the heating up effect we proposed a channel model
where the variance of the additive noise depends on a weighted sum of
the past channel input powers. The weights characterize the efficiency
of the heat sink.

To study the capacity of this channel at low SNR, we
computed the capacity per unit cost. We showed that the heating
effect is not just unharmful but can be even beneficial in the sense
that the capacity per unit cost
can be larger than the capacity per unit cost of a corresponding
channel with ideal heat sink, i.e., where the weights describing the
dependency of the noise variance on the channel input powers are
zero. This suggests that at low SNR no heat sinks should be used.

Studying capacity at high SNR, we derived a sufficient condition and a
necessary condition on the weights for the capacity to be bounded in
the SNR. We showed that when the sequence of weights decays not faster than
geometrically, then capacity is bounded in the SNR. On the other hand,
if the sequence of weights decays faster than geometrically, then capacity is
unbounded in the SNR. This result demonstrates the importance of an
efficient heat sink at high SNR.

\section*{Acknowledgment}
Fruitful discussions with Ashish Khisti and Mich\`ele Wigger are
gratefully acknowledged. Sergio Verd\'u's comments at the ISIT 2007 on
our low SNR results are also much appreciated.

\appendix

\section{Proof of Proposition~\ref{prop:lowSNR}}
\label{app:proplowSNR}
We first note that by the expression of the capacity per unit cost of a memoryless
channel \cite{verdu90} we have
\begin{equation}
  \sup_{\SNR>0} \frac{C_{\alpha=0}(\SNR)}{\SNR} = \sup_{\zeta^2 > 0}
  \frac{D\big(W_{\alpha=0}(\cdot|\zeta)\big\|
  W_{\alpha=0}(\cdot|0)\big)}{\zeta^2/\sigma^2}, \label{eq:unitcostmemoryless}
\end{equation}
where $W_{\alpha=0}(\cdot|\cdot)$ denotes the channel law of the
channel
\begin{equation}
  Y_k = x_k + \sigma \cdot U_k. \label{eq:appmemoryless}
\end{equation}
Thus to prove Proposition~\ref{prop:lowSNR} it suffices to show that
\begin{equation*}
  \sup_{\SNR>0} \frac{C_{\textnormal{Info}}(\SNR)}{\SNR} \geq \sup_{\zeta^2 > 0}
  \frac{D\big(W_{\alpha=0}(\cdot|\zeta)\big\|
  W_{\alpha=0}(\cdot|0)\big)}{\zeta^2/\sigma^2}.
\end{equation*}
We shall obtain this result by deriving a lower bound on
$C_{\textnormal{Info}}(\SNR)$ and by computing then its limiting ratio to
$\SNR$ as $\SNR$ tends to zero.

In order to lower bound $C_{\textnormal{Info}}(\SNR)$, which was
defined in \eqref{eq:info} as
\begin{equation*}
  C_{\textnormal{Info}}(\SNR) = \varliminf_{n \to \infty} \frac{1}{n} \sup I(X_1^n;Y_1^n),
\end{equation*}
we evaluate $\frac{1}{n} I(X_1^n;Y_1^n)$ for inputs $\{X_k\}$ that are
blockwise IID in blocks of $L$ symbols (for some $L\in\Integers^+$). Thus
$\{(X_{bL+1},\ldots,X_{(b+1)L}),b\in\Integers_0^+\}$ is a sequence of
IID random length-$L$ vectors with $(X_{bL+1},\ldots,X_{(b+1)L})$ taking on the value
$(\xi,0,\ldots,0)$ with probability $\delta$ and $(0,\ldots,0)$ with probability
$1-\delta$, for some $\xi \in \Reals$. To satisfy the power constraint
\eqref{eq:power} we shall choose $\xi$ and $\delta$ such that
\begin{equation}
  \frac{\xi^2}{\sigma^2}\delta 
  = L \: \SNR. \label{eq:apppower}
\end{equation}

We use the chain rule for mutual information to write
\begin{IEEEeqnarray}{lCl}
  \frac{1}{n} I(X_1^n;Y_1^n) & = & \frac{1}{n} \sum_{b=0}^{\lfloor n/L \rfloor -1}
  I\big(X_{b L+1};Y_1^n\big|X_1^{b L}\big) \nonumber\\
  & \geq & \frac{1}{n} \sum_{b=0}^{\lfloor n/L \rfloor -1}
  I\big(X_{bL+1};Y_{bL+1}\big|X_1^{bL}\big), \label{eq:app1}
\end{IEEEeqnarray}
where the inequality follows because reducing
observations cannot increase mutual information.

Let $R_{\textnormal{on-off}}^{(\xi)}(\textsf{snr})$ denote the maximum rate achievable
on \eqref{eq:appmemoryless} using on-off keying
with on-symbol $\xi$ and with its corresponding probability
$\wp$ chosen in order to satisfy the power constraint $\textsf{snr}$,
i.e.,
\begin{equation}
  R_{\textnormal{on-off}}^{(\xi)}\left(\textsf{snr}\right) \triangleq
  \sup_{\substack{P_X(\xi)=1-P_X(0)=\wp,\\\xi^2/\sigma^2
  \wp \leq \textsf{snr}}} I(X;X+\sigma\cdot U_k), \qquad \textsf{snr}\geq 0.\label{eq:appdefinerate}
\end{equation}
Notice that $R^{(\xi)}_{\textnormal{on-off}}(\textsf{snr})$, $\textsf{snr}\geq 0$ is a
nonnegative, monotonically nondecreasing function
of $\textsf{snr}$ with $R^{(\xi)}_{\textnormal{on-off}}(0)=0$. From the strict
concavity of mutual information it follows that
$R^{(\xi)}_{\textnormal{on-off}}(\textsf{snr})>0$ whenever $\textsf{snr}>0$. Also, for a
fixed $\xi$, $\textsf{snr} \mapsto R^{(\xi)}_{\textnormal{on-off}}(\textsf{snr})$ is concave in
$\textsf{snr}$. Consequently, for some $\textsf{snr}_0>0$, the function
$\textsf{snr} \mapsto R^{(\xi)}_{\textnormal{on-off}}(\textsf{snr})$ is strictly monotonic in the
interval $\textsf{snr}\in[0,\textsf{snr}_0]$, and hence the supremum on the RHS of
\eqref{eq:appdefinerate} is attained for
$\wp=\textsf{snr}\:\sigma^2/\xi^2$, $\textsf{snr}\in[0,\textsf{snr}_0]$.

By writing $I(X_{bL+1};Y_{bL+1}|X_1^{bL}=x_1^{bL})$ for a given
 $X_1^{bL}=x_1^{bL}$ as
 \begin{IEEEeqnarray*}{lCl}
   I\big(X_{bL+1};Y_{bL+1}\big|X_1^{bL}=x_1^{bL}\big) 
   & = & I\left(X_{bL+1};X_{bL+1}+\theta\big(x_1^{bL}\big)\cdot U_{bL+1}\right) \\
   & = & I\left(X_{bL+1};\frac{\sigma}{\theta\big(x_1^{bL}\big)}X_{bL+1}+\sigma \cdot U_{bL+1}\right)
 \end{IEEEeqnarray*}
(with $\theta\big(x_1^{bL}\big)$ defined in \eqref{eq:sigma}), and by 
using that for $\textsf{snr}\in[0,\textsf{snr}_0]$ the supremum on the RHS of \eqref{eq:appdefinerate} is
attained for $\wp=\textsf{snr}\:\sigma^2/\xi^2$ we obtain
\begin{IEEEeqnarray}{lCl}
  I\big(X_{bL+1};Y_{bL+1}\big|
    X_1^{bL}=x_1^{bL}\big)
    & = &  
  R^{(\xi)}_{\textnormal{on-off}}\Bigg(\frac{L\:\SNR}{1+\sum_{\ell=0}^{b-1}\alpha_{(b-\ell)
        L}x_{\ell L+1}^2/\sigma^2}\Bigg), \;\;\; \SNR\in[0,\SNR_0],\IEEEeqnarraynumspace
\end{IEEEeqnarray}
where $\SNR_0\triangleq\textsf{snr}_0/L$. Averaging over $X_1^{bL}$ and
combining with \eqref{eq:app1} yields
\begin{IEEEeqnarray}{lCl}
  \frac{1}{n} I(X_1^n;Y_1^n) & \geq & \frac{1}{n} \sum_{b=0}^{\lfloor
    n/L \rfloor -1}
  \E{R^{(\xi)}_{\textnormal{on-off}}\left(\frac{L\:\SNR}{1+\sum_{\ell=0}^{b-1}\alpha_{(b-\ell)L}X_{\ell
          L+1}^2/\sigma^2}\right)}\nonumber\\
  & \geq & \frac{\lfloor n/L \rfloor}{n}
    R^{(\xi)}_{\textnormal{on-off}}\left(\frac{L\:
    \SNR}{1+\sum_{\ell=1}^{\infty} \alpha_{\ell
    L}\xi^2/\sigma^2}\right), \quad \SNR\in[0,\SNR_0],\IEEEeqnarraynumspace
\end{IEEEeqnarray}
where the second inequality follows by upper bounding
$\sum_{\ell=0}^{b-1}\alpha_{(b-\ell)L}X_{\ell L+1}^2/\sigma^2 \leq
\sum_{\ell=1}^{\infty} \alpha_{\ell L} \xi^2/\sigma^2$, and by using
that $\textsf{snr}\mapsto R^{(\xi)}_{\textnormal{on-off}}(\textsf{snr})$ is monotonically
increasing in $\textsf{snr}$. The lower bound on $C_{\textnormal{Info}}(\SNR)$
follows then by letting $n$ tend to infinity
\begin{equation}
  C_{\textnormal{Info}}(\SNR) = \varliminf_{n \to \infty} \frac{1}{n} I(X_1^n;Y_1^n) \geq
  \frac{1}{L}  R^{(\xi)}_{\textnormal{on-off}}\left(\frac{L\:
      \SNR}{1+\sum_{\ell=1}^{\infty} \alpha_{\ell
        L}\xi^2/\sigma^2}\right). \label{eq:appLB1}
\end{equation}
With this we can lower bound the information capacity per unit cost as
\begin{IEEEeqnarray}{lCl}
  \sup_{\SNR>0} \frac{C_{\textnormal{Info}}(\SNR)}{\SNR} & \geq &
  \lim_{\SNR\downarrow 0} \frac{C_{\textnormal{Info}}(\SNR)}{\SNR} \nonumber\\
  & \geq &
  \lim_{\SNR\downarrow 0} \frac{1}{L} \frac{ R^{(\xi)}_{\textnormal{on-off}}\left(\frac{L\:
      \SNR}{1+\sum_{\ell=1}^{\infty} \alpha_{\ell
        L}\xi^2/\sigma^2}\right)}{\SNR} \nonumber\\
  & = & \lim_{\SNR\downarrow 0}  \frac{R^{(\xi)}_{\textnormal{on-off}}\left(\frac{L\:
        \SNR}{1+\sum_{\ell=1}^{\infty} \alpha_{\ell
          L}\xi^2/\sigma^2}\right)}{\frac{L\:\SNR}{1+\sum_{\ell=1}^{\infty}\alpha_{\ell L}\xi^2/\sigma^2}} \: \frac{1}{1+\sum_{\ell=1}^{\infty} \alpha_{\ell L} \xi^2/\sigma^2} \nonumber\\
  & = & \lim_{\SNR'\downarrow 0}
  \frac{R^{(\xi)}_{\textnormal{on-off}}(\SNR')}{\SNR'} \:
  \frac{1}{1+\sum_{\ell=1}^{\infty} \alpha_{\ell L}
    \xi^2/\sigma^2},\nonumber\\
\end{IEEEeqnarray}
where the first inequality follows by lower bounding the supremum by
  the limit; and where the last equality follows by substituting
  \mbox{$\SNR' =
  \frac{L\:\SNR}{1+\sum_{\ell=1}^{\infty}\alpha_{\ell
  L}\xi^2/\sigma^2}$}.

Proceeding along the lines of the proof of
\cite[Thm.~3]{verdu90}, it can be shown that
\begin{equation}
  \lim_{\SNR'\downarrow 0} \frac{R^{(\xi)}_{\textnormal{on-off}}(\SNR')}{\SNR'} = \frac{D\big(W_{\alpha=0}(\cdot|\xi)\big\|W_{\alpha=0}(\cdot|0)\big)}{\xi^2/\sigma^2}
\end{equation}
and therefore
\begin{equation}
  \sup_{\SNR > 0} \frac{C_{\textnormal{Info}}(\SNR)}{\SNR} \geq
  \frac{D\big(W_{\alpha=0}(\cdot|\xi)\big\|W_{\alpha=0}(\cdot|0)\big)}{\xi^2/\sigma^2}
  \cdot \frac{1}{1+\sum_{\ell=1}^{\infty} \alpha_{\ell L} \xi^2/\sigma^2}.
\end{equation}
Noting that \eqref{eq:bounded} \& \eqref{eq:alpha} imply
\begin{equation}
  0 \leq \lim_{L \to \infty} \sum_{\ell=1}^{\infty} \alpha_{\ell L} \leq
  \lim_{L \to \infty} \sum_{\ell=L}^{\infty} \alpha_{\ell} = 0 \label{eq:appalphaL}
\end{equation}
we obtain by letting $L$ tend to infinity
\begin{equation}
  \sup_{\SNR>0} \frac{C_{\textnormal{Info}}(\SNR)}{\SNR} \geq
  \frac{D\big(W_{\alpha=0}(\cdot|\xi)\big\|W_{\alpha=0}(\cdot|0)\big)}{\xi^2/\sigma^2}. \label{eq:appbeforemax}
\end{equation}
Maximizing \eqref{eq:appbeforemax} over $\xi^2$ yields then
\begin{equation}
  \sup_{\SNR>0} \frac{C_{\textnormal{Info}}(\SNR)}{\SNR} \geq
  \sup_{\xi^2>0}
  \frac{D\big(W_{\alpha=0}(\cdot|\xi)\big\|W_{\alpha=0}(\cdot|0)\big)}{\xi^2/\sigma^2}
\end{equation}
which, in view of \eqref{eq:unitcostmemoryless}, proves Proposition~\ref{prop:lowSNR}.

\section{Appendix to Section~\ref{sub:lowerLOW}}
\label{app:tozero}
We shall prove that
\begin{equation}
  \lim_{b \to \infty} I\big(\vect{X}_{-\infty}^{-1};\tilde{\vect{Y}}_{b}\big|\vect{X}_0^{b}\big)
  = 0.\label{eq:tozero}
\end{equation}
Let $\alpha_{b}^{(i)}$ be defined as
\begin{IEEEeqnarray}{lCl}
  \alpha^{(1)}_0 & \triangleq & 0 \\
  \alpha_{b}^{(i)} & \triangleq & \alpha_{b L+i-1}, \qquad
  (b,i) \in \Integers^+_0 \times \Integers^+\setminus\{(0,1)\}.
\end{IEEEeqnarray}
We have
\begin{IEEEeqnarray}{lCl}
  I\big(\vect{X}_{-\infty}^{-1};\tilde{\vect{Y}}_{b}\big|\vect{X}_0^{b}\big)
  & = & \sum_{i=1}^L I\big(\vect{X}_{-\infty}^{-1};\tilde{Y}_{b L +
  i}\big|\vect{X}_0^{b}, \tilde{Y}_{b L+1}^{b L+i-1}\big)
  \nonumber\\
  & \leq & \sum_{i=1}^L \Bigl(h\big(\tilde{Y}_{bL+i}\big|\vect{X}_0^b\big)-h\big(\tilde{Y}_{bL+i}\big|\vect{X}_{-\infty}^b\big)\Bigr)\nonumber\\
  & \leq & \frac{1}{2}\sum_{i=1}^L \E{\log\left((2\pi
  e)\left(\sigma^2+\sum_{\ell=0}^{b}\alpha_{b-\ell}^{(i)}X_{\ell L+1}^2+\const{P}\:
  L \sum_{\ell=b+1}^{\infty} \alpha_{\ell}^{(i)}\right)\right)}
  \nonumber\\
  & & {} - \frac{1}{2} \sum_{i=1}^L \E{\log\left((2\pi e)
      \left(\sigma^2+\sum_{\ell=0}^{b} \alpha_{b-\ell}^{(i)}
        X_{\ell L+1}^2+\sum_{\ell=-\infty}^{-1}\alpha_{b-\ell}^{(i)}
        X_{\ell L+1}^2\right)\right)}\nonumber\\
  & \leq & \frac{1}{2}\sum_{i=1}^L \E{\log\left((2\pi
      e)\left(\sigma^2+\sum_{\ell=0}^{b}\alpha_{b-\ell}^{(i)}X_{\ell L+1}^2+\const{P}\:
        L \sum_{\ell=b+1}^{\infty} \alpha_{\ell}^{(i)}\right)\right)}
  \nonumber\\
  & & {} - \frac{1}{2} \sum_{i=1}^L \E{\log\left((2\pi e)
      \left(\sigma^2+\sum_{\ell=0}^{b} \alpha_{b-\ell}^{(i)}
        X_{\ell L+1}^2\right)\right)}\nonumber\\
  & = & \frac{1}{2} \sum_{i=1}^L \E{\log\left(1+\frac{\const{P}\: L
        \sum_{\ell=b+1}^{\infty} \alpha_{\ell}^{(i)}}{\sigma^2 +
        \sum_{\ell=0}^{b} \alpha_{b-\ell}^{(i)}
        X_{\ell L+1}^2}\right)}\nonumber\\
  & \leq & \frac{1}{2}\sum_{i=1}^L \log\left(1+ L\:\SNR
  \sum_{\ell=b+1}^{\infty} \alpha_{\ell}^{(i)}\right),
\end{IEEEeqnarray}
where the first inequality follows because conditioning cannot
increase entropy and because, conditional on $\vect{X}_{-\infty}^b$,
$\tilde{Y}_{bL+i}$ is independent of $\tilde{Y}_{bL+1}^{bL+i-1}$; the next inequality follows from the entropy maximizing
property of Gaussian random variables; the subsequent inequality
follows because
$\sum_{\ell=-\infty}^{-1} \alpha_{b-\ell}^{(i)} X_{\ell L+1}^2 \geq
0$, $i=1,\ldots,L$; and the last
inequality follows because
$\sum_{\ell=0}^{b} \alpha_{b-\ell}^{(i)} X_{\ell L+1}^2 \geq 0$, $i=1,\ldots,L$.

By upper bounding
\begin{equation}
  \sum_{\ell=b+1}^{\infty} \alpha_{\ell}^{(i)} \leq
  \sum_{\ell=b+1}^{\infty} \alpha_{\ell}, \qquad i=1,\ldots,L
\end{equation}
we obtain
\begin{equation}
  I\big(\vect{X}_{-\infty}^{-1};\tilde{\vect{Y}}_{b}\big|\vect{X}_0^{b}\big)
  \leq \frac{L}{2} \log\left(1+L\:\SNR \sum_{\ell=b+1}^{\infty} \alpha_{\ell}\right),
\end{equation}
and \eqref{eq:tozero} follows by noting that \eqref{eq:alpha}
implies
\begin{equation*}
  \lim_{b \to \infty} \sum_{\ell=b+1}^{\infty} \alpha_{i} = 0.
\end{equation*}

\section{Proof of Lemma~\ref{lemma:typical}}
\label{app:lemmatypical}
We shall show that for any $\eps>0$
\begin{equation}
  \lim_{n\to \infty} \Prob\left(\left|\frac{1}{\lfloor n/L \rfloor}
  \|\vect{Y}\|^2-(\sigma^2+\const{P}+\alpha^{(L)}\:
  \const{P})\right|\geq \eps\right) = 0 \label{eq:typicalP1}
\end{equation}
and
\begin{equation}
  \lim_{n \to \infty} \Prob\left(\left|\frac{1}{\lfloor n/L \rfloor}
  \|\vect{Z}\|^2-(\sigma^2+\alpha^{(L)}\:
  \const{P})\right|\geq \eps\right) = 0.\label{eq:typicalP2}
\end{equation}
Lemma~\ref{lemma:typical} follows then by the union of events
bound.

In order to prove \eqref{eq:typicalP1} \& \eqref{eq:typicalP2},
we first note that
\begin{IEEEeqnarray}{lCl}
  \frac{1}{\lfloor n/L \rfloor} \E{\|\vect{Y}\|^2} & = & \sigma^2 +
  \const{P} + \frac{\const{P}}{\lfloor n/L \rfloor}
  \sum_{k=1}^{\lfloor n/L \rfloor -1} \sum_{\ell=1}^{k}
  \alpha_{\ell L} \label{eq:typicalmean}\\
  \frac{1}{\lfloor n/L \rfloor} \E{\|\vect{Z}\|^2} & = & \sigma^2 +
  \frac{\const{P}}{\lfloor n/L \rfloor} \sum_{k=1}^{\lfloor n/L
  \rfloor-1} \sum_{\ell=1}^{k} \alpha_{\ell L}
\end{IEEEeqnarray}
and therefore, by Ces\'aro's mean \cite[Thm.~4.2.3]{coverthomas91},
\begin{IEEEeqnarray}{lCl}
  \lim_{n \to \infty} \frac{1}{\lfloor n/L \rfloor} \E{\|\vect{Y}\|^2}
  & = & \sigma^2 + \const{P} + \alpha^{(L)}\:\const{P}\\
  \lim_{n \to \infty} \frac{1}{\lfloor n/L \rfloor} \E{\|\vect{Z}\|^2}
  & = & \sigma^2 + \alpha^{(L)}\:\const{P},
\end{IEEEeqnarray}
where $\alpha^{(L)}$ was defined in \eqref{eq:lemmaalphaL} as
\begin{equation*}
  \alpha^{(L)} = \sum_{\ell=1}^{\infty} \alpha_{\ell L}.
\end{equation*}
Thus, for any $\eps>0$ and $0<\varepsilon<\eps$, there exists an $n_0$
such that for all $n\geq n_0$
\begin{IEEEeqnarray}{cCl}
  \left|\frac{1}{\lfloor n/L \rfloor} \E{\|\vect{Y}\|^2} -
    (\sigma^2 + \const{P}+\alpha^{(L)}\:\const{P}) \right| & \leq &
  \varepsilon \\
  \left|\frac{1}{\lfloor n/L \rfloor} \E{\|\vect{Z}\|^2} -
    (\sigma^2 +\alpha^{(L)}\:\const{P}) \right| & \leq &
  \varepsilon
\end{IEEEeqnarray}
and it follows from the triangle inequality that
\begin{IEEEeqnarray}{cCl}
  \left|\frac{1}{\lfloor n/L \rfloor} \|\vect{Y}\|^2
   -(\sigma^2+\const{P}+\alpha^{(L)}\:
    \const{P})\right| & \leq & \left|\frac{1}{\lfloor n/L \rfloor}\|\vect{Y}\|^2
    -\frac{1}{\lfloor n/L \rfloor}\E{\|\vect{Y}\|^2}
    \right| + \varepsilon \IEEEeqnarraynumspace\\
  \left|\frac{1}{\lfloor n/L \rfloor}\|\vect{Z}\|^2
    -(\sigma^2+\alpha^{(L)}\:
    \const{P})\right| & \leq & \left|\frac{1}{\lfloor n/L \rfloor}\|\vect{Z}\|^2
    -\frac{1}{\lfloor n/L \rfloor}\E{\|\vect{Z}\|^2}
    \right| + \varepsilon.
\end{IEEEeqnarray}
From this we obtain
\begin{IEEEeqnarray}{lCl}
  \Prob\left(\left|\frac{1}{\lfloor n/L \rfloor}
        \|\vect{Y}\|^2-(\sigma^2+\const{P}+\alpha^{(L)}\:
        \const{P})\right|\geq \eps\right)
  & \leq & \Prob\left(\left|\frac{1}{\lfloor n/L \rfloor}
      \|\vect{Y}\|^2-\frac{1}{\lfloor n/L
        \rfloor}\E{\|\vect{Y}\|^2}\right|\geq
    \eps-\varepsilon\right)\nonumber\\
  & \leq & \frac{\Var{\frac{1}{\lfloor n/L
        \rfloor}\|\vect{Y}\|^2}}{(\eps-\varepsilon)^2}
        \label{eq:typicalcheby1}
\end{IEEEeqnarray}
and
\begin{IEEEeqnarray}{lCl}
  \Prob\left(\left|\frac{1}{\lfloor n/L \rfloor}
        \|\vect{Z}\|^2-(\sigma^2+\alpha^{(L)}\:
        \const{P})\right|\geq \eps\right)
 & \leq & \Prob\left(\left|\frac{1}{\lfloor n/L \rfloor}
      \|\vect{Z}\|^2-\frac{1}{\lfloor n/L
        \rfloor}\E{\|\vect{Z}\|^2}\right|\geq
    \eps-\varepsilon\right)\nonumber\\
  & \leq & \frac{\Var{\frac{1}{\lfloor n/L
        \rfloor}\|\vect{Z}\|^2}}{(\eps-\varepsilon)^2},
        \label{eq:typicalcheby2}
\end{IEEEeqnarray}
with $\Var{A}=\E{(A-\E{A})^2}$ denoting the variance of $A$. Here the
last inequalities in \eqref{eq:typicalcheby1} \&
\eqref{eq:typicalcheby2} follow from Chebyshev's inequality
\cite[Sec.~5.4]{gallager68}.

It remains to show that
\begin{equation}
  \lim_{n \to \infty} \Var{\frac{1}{\lfloor n/L
  \rfloor}\|\vect{Y}\|^2} = \lim_{n \to \infty} \Var{\frac{1}{\lfloor
  n/L \rfloor} \| \vect{Z}\|^2} = 0. \label{eq:typicalzerovar}
\end{equation}
We shall prove \eqref{eq:typicalzerovar} for $\vect{Y}$. The proof for
$\vect{Z}$ follows along the same lines. We begin by writing
$\Var{\frac{1}{\lfloor n/L \rfloor}\|\vect{Y}\|^2}$ as
\begin{IEEEeqnarray}{lCl}
  \IEEEeqnarraymulticol{3}{l}{\Var{\frac{1}{\lfloor n/L
        \rfloor}\|\vect{Y}\|^2}}\nonumber\\
  \qquad & = &
  \frac{1}{\big(\lfloor n/L \rfloor\big)^2} \Var{\sum_{k=0}^{\lfloor n/L
      \rfloor -1} Y_{kL+1}^2} \nonumber\\
  & = & \frac{1}{\big(\lfloor n/L \rfloor\big)^2}\sum_{k=0}^{\lfloor
  n/L \rfloor -1} \Var{Y_{kL+1}^2} + \frac{2}{\big(\lfloor n/L
  \rfloor\big)^2} \sum_{\substack{k=1, j=0\\k > j}}^{\lfloor n/L \rfloor -1}
        \Co{Y_{kL+1}^2,Y_{jL+1}^2}, \IEEEeqnarraynumspace \label{eq:typicalvar}
\end{IEEEeqnarray}
where $\Co{A,B}=\E{(A-\E{A})(B-\E{B})}$ denotes the covariance between
$A$ and $B$. We shall evaluate both terms on the RHS of
\eqref{eq:typicalvar} separately. For the
sake of clarity, we shall omit the details of the derivations and
show only the main steps. Unless otherwise stated these steps can be
derived in a straightforward way using that
\begin{enumerate}
  \renewcommand{\labelenumi}{\roman{enumi})}
\item $\{X_{kL+1}\, ,\, k\in\Integers_0^+\}$ is a sequence of
  IID, zero-mean, variance-$\const{P}$ Gaussian random variables whose
  fourth moments are given by $3\const{P}$, while all odd moments are
  zero;
\item $X_k=0$ if $k \mod L \neq 1$;
\item $\{U_{k}\}$ (and hence also $\{U_{kL+1}\, ,\,
  k\in\Integers_0^+\}$) is a zero-mean, unit-variance, stationary \&
  weakly-mixing random process;
\item and that $\{X_k\}$ and $\{U_k\}$ are independent of each other.
\end{enumerate}

For the first sum on the RHS of \eqref{eq:typicalvar} it suffices to
show that $\Var{Y_{kL+1}}<\infty$, $k\in\Integers^+_0$. Indeed,
this sum
contains only
$\lfloor n/L \rfloor$ summands and hence, when divided by $(\lfloor n/L \rfloor)^2$, this
sum vanishes as $n$ tends to
infinity, given that $\Var{Y_{kL+1}}<\infty$, $k\in\Integers^+_0$. We have
\begin{IEEEeqnarray}{lCl}
  \Var{Y_{kL+1}^2}
  & = & \E{Y_{kL+1}^4}-\left(\E{Y_{kL+1}^2}\right)^2\nonumber\\
  & \leq & \E{Y_{kL+1}^4} \nonumber\\
  & = & \E{\left(X_{kL+1}+\theta\big(X_1^{kL}\big)\cdot U_{kL+1}\right)^4} \nonumber\\
  & = & 3 \const{P}^2 + 6
  \const{P}\left(\sigma^2+\const{P}\sum_{\ell=1}^{k} \alpha_{\ell
      L}\right) \nonumber\\
  && {} +
  \left(\sigma^4+2\sigma^2\const{P}\sum_{\ell=1}^{k}\alpha_{\ell L}
  +2\const{P}^2 \sum_{\ell=1}^{k} \alpha_{\ell L}^2 + \const{P}^2
  \left(\sum_{\ell=1}^{k} \alpha_{kL}\right)^2\right)\E{U_{kL+1}^4}
  \nonumber\\
  & \leq & 3 \const{P}^2 +
  6\const{P}\left(\sigma^2+\const{P}
    \alpha^{(L)}\right) \nonumber\\
  & & {} + \left(\sigma^4+2\sigma^2\const{P} \alpha^{(L)}+2
    \const{P}^2 \sum_{\ell=1}^{\infty} \alpha_{\ell L}^2
    +\const{P}^2\left(\alpha^{(L)}\right)^2\right)\E{U_{kL+1}^4}
\end{IEEEeqnarray}
where the second inequality follows by upper bounding
$\sum_{\ell=1}^k\alpha_{\ell L}\leq\alpha^{(L)}$. Note that \eqref{eq:positivecoefficients} implies
that $\alpha^{(L)}$ and $\sum_{\ell=1}^{\infty} \alpha_{\ell L}^2$ are
bounded. It follows therefore by noting that $U_{kL+1}$ has a finite
fourth moment that (for a finite $\const{P}$) 
\begin{equation}
\Var{Y_{kL+1}} < \infty. \label{eq:typicalfinite}
\end{equation}

In order to show that the second term on the RHS of
\eqref{eq:typicalvar} vanishes as $n$ tends to infinity, we shall
evaluate 
\begin{equation*}
\Co{Y_{kL+1},Y_{jL+1}} = \E{Y_{kL+1}^2Y_{jL+1}^2}-\E{Y_{kL+1}^2}\E{Y_{jL+1}^2}
\end{equation*}
for $k\in\Integers^+$,
$j\in\Integers^+_0$, $k>j$. We have
\begin{IEEEeqnarray}{lCl}
  \E{Y_{kL+1}^2 Y_{jL+1}^2}
  & = & \E{\left(X_{kL+1}+\theta\big(X_1^{kL}\big)\cdot
      U_{kL+1}\right)^2\left(X_{jL+1}+\theta\big(X_1^{jL}\big)\cdot
      U_{jL+1}\right)^2}\nonumber\\
  & = & \const{P}^2 +\const{P}
  \left(\sigma^2+\const{P}\sum_{\ell=1}^{j}\alpha_{\ell L}\right) +
  \const{P}\left(\sigma^2+\const{P}\sum_{\ell=1}^{k}\alpha_{\ell
      L}\right) + 2 \const{P}^2\alpha_{(k-j)L} \nonumber\\
  & & {} + \left(\sigma^2+\const{P}\sum_{\ell=1}^{k}\alpha_{\ell
      L}\right)\left(\sigma^2+\const{P}\sum_{\ell'=1}^{j}\alpha_{\ell'
      L}\right)\E{U_{kL+1}^2U_{jL+1}^2}\nonumber\\
  & & {} +2\const{P}^2\sum_{\ell=1}^{j} \alpha_{\ell
  L}\alpha_{(\ell+k-j)L}\: \E{U_{kL+1}^2U_{jL+1}^2}.\label{eq:typicalcorr}
\end{IEEEeqnarray}
Evaluating
\begin{IEEEeqnarray}{lCl}
  \E{Y_{kL+1}^2}\E{Y_{jL+1}^2} 
  & = &
  \const{P}^2+\const{P}\left(\sigma^2+\const{P}\sum_{\ell=1}^{j}\alpha_{\ell
      L}\right) + \const{P}\left(\sigma^2+\const{P}\sum_{\ell=1}^{k}\alpha_{\ell
      L}\right) \nonumber\\
  && {} + \left(\sigma^2+\const{P}\sum_{\ell=1}^{k}\alpha_{\ell
  L}\right)\left(\sigma^2+\const{P}\sum_{\ell'=1}^{j}\alpha_{\ell'
  L}\right) \label{eq:typicalproductmean}
\end{IEEEeqnarray}
we obtain from \eqref{eq:typicalproductmean} \& \eqref{eq:typicalcorr}
\begin{IEEEeqnarray}{lCl}
  \Co{Y_{kL+1},Y_{jL+1}}
  & = & 2\const{P}^2
  \alpha_{(k-j)L}+2\const{P}^2\sum_{\ell=1}^{j}\alpha_{\ell L}
  \alpha_{(\ell+k-j)L}\E{U_{kL+1}^2U_{jL+1}^2} \nonumber\\
  & & {}+ \left(\sigma^2+\const{P}\sum_{\ell=1}^{k}\alpha_{\ell
  L}\right)\left(\sigma^2+\const{P}\sum_{\ell'=1}^{j}\alpha_{\ell'
  L}\right)\left(\E{U_{kL+1}^2U_{jL+1}^2}-1\right).\IEEEeqnarraynumspace
\end{IEEEeqnarray}
Summing over $k$ and $j$ and diving by $(\lfloor n/L\rfloor)^2$ yields
\begin{IEEEeqnarray}{lCl}
  \IEEEeqnarraymulticol{3}{l}{\frac{2}{(\lfloor n/L \rfloor)^2} \sum_{\substack{k=1, j=0\\k > j}}^{\lfloor n/L \rfloor -1}
    \Co{Y_{kL+1}^2,Y_{jL+1}^2}}\nonumber\\
  \quad & = & \frac{2}{(\lfloor n/L \rfloor)^2} \sum_{\substack{k=1, j=0\\k
      > j}}^{\lfloor n/L \rfloor -1} \Biggl(2\const{P}^2
    \alpha_{(k-j)L}+2\const{P}^2\sum_{\ell=1}^{j}\alpha_{\ell L}
    \alpha_{(\ell+k-j)L}\E{U_{kL+1}^2U_{jL+1}^2} \nonumber\\
  & & {} \qquad \qquad \qquad \qquad \, + \left(\sigma^2+\const{P}\sum_{\ell=1}^{k}\alpha_{\ell
        L}\right)\left(\sigma^2+\const{P}\sum_{\ell'=1}^{j}\alpha_{\ell'
        L}\right)\left(\E{U_{kL+1}^2U_{jL+1}^2}-1\right)\Biggr)\nonumber\\
  & = & \frac{2}{(\lfloor n/L \rfloor)^2} \sum_{j=0}^{\lfloor n/L
    \rfloor -2} \sum_{\nu=1}^{\lfloor n/L \rfloor-1-j} \Biggl(2\const{P}^2
    \alpha_{\nu L}+2\const{P}^2\sum_{\ell=1}^{j}\alpha_{\ell L}
    \alpha_{(\ell+\nu)L}\E{U_{\nu L+1}^2U_{1}^2}\nonumber\\
  & & {} \qquad \qquad \qquad \qquad \qquad \qquad\,\, + \left(\sigma^2+\const{P}\sum_{\ell=1}^{j+\nu}\alpha_{\ell
        L}\right)\left(\sigma^2+\const{P}\sum_{\ell'=1}^{j}\alpha_{\ell'
        L}\right)\left(\E{U_{\nu L+1}^2U_{1}^2}-1\right)\Biggr)
    \nonumber\\
    & = & \frac{2}{(\lfloor n/L \rfloor)^2} \sum_{j=0}^{\lfloor n/L
    \rfloor -2} \sum_{\nu=1}^{\lfloor n/L \rfloor-1-j} 2\const{P}^2
    \alpha_{\nu L} \nonumber\\
    & & {}+ \frac{2}{(\lfloor n/L \rfloor)^2} \sum_{j=0}^{\lfloor n/L
    \rfloor -2} \sum_{\nu=1}^{\lfloor n/L \rfloor-1-j} 2\const{P}^2\sum_{\ell=1}^{j}\alpha_{\ell L}
    \alpha_{(\ell+\nu)L}\E{U_{\nu L+1}^2U_{1}^2}\nonumber\\
  & & {} + \frac{2}{(\lfloor n/L \rfloor)^2} \sum_{j=0}^{\lfloor n/L
    \rfloor -2} \sum_{\nu=1}^{\lfloor n/L \rfloor-1-j}\left(\sigma^2+\const{P}\sum_{\ell=1}^{j+\nu}\alpha_{\ell
        L}\right)\left(\sigma^2+\const{P}\sum_{\ell'=1}^{j}\alpha_{\ell'
        L}\right)\left(\E{U_{\nu L+1}^2U_{1}^2}-1\right),\;\;\nonumber\\
 \label{eq:typicalsum1}
\end{IEEEeqnarray}
where the second equality follows by substituting $\nu=k-j$ and from
the stationarity of $\{U_k\}$. 

The first
two terms on the RHS of \eqref{eq:typicalsum1} can be upper bounded
using \eqref{eq:positivecoefficients}
\begin{equation*}
  \alpha_{\ell} < \varrho^{\ell}, \qquad 0<\varrho<1, \quad \ell\geq\ell_0.
\end{equation*}
Indeed, noting that $L \geq \ell_0$, we have
\begin{equation}
  \sum_{\nu=1}^{\lfloor n/L
  \rfloor-1-j} \alpha_{\nu L} < \sum_{\nu=1}^{\lfloor n/L
  \rfloor-1-j} \varrho^{\nu L} < \sum_{\nu=1}^{\lfloor n/L
  \rfloor} \varrho^{\nu L} \label{eq:typicaldelta1}
\end{equation}
and
\begin{IEEEeqnarray}{lCl}
  \sum_{\nu=1}^{\lfloor n/L
  \rfloor-1-j} \sum_{\ell=1}^{j}\alpha_{\ell L}
    \alpha_{(\ell+\nu)L} & < & \sum_{\nu=1}^{\lfloor n/L
  \rfloor-1-j} \sum_{\ell=1}^{j} \left(\varrho^{2 L}\right)^{\ell}
  \varrho^{\nu L} \nonumber\\
  & < &  \sum_{\nu=1}^{\lfloor n/L
  \rfloor} \sum_{\ell=1}^{\infty} \left(\varrho^{2 L}\right)^{\ell}
  \varrho^{\nu L}\nonumber\\
  & = & \frac{\varrho^{2L}}{1-\varrho^{2L}} \sum_{\nu=1}^{\lfloor n/L
  \rfloor} \varrho^{\nu L}.\label{eq:typicaldelta2}
\end{IEEEeqnarray}
Consequently with \eqref{eq:typicaldelta1} we can upper bound the first term on the RHS of
\eqref{eq:typicalsum1} as
\begin{IEEEeqnarray}{lCl}
  \frac{2}{(\lfloor n/L \rfloor)^2}\sum_{j=0}^{\lfloor n/L
      \rfloor -2} \sum_{\nu=1}^{\lfloor n/L \rfloor-1-j} 2
    \const{P}^2 \alpha_{\nu L}
  & < & \frac{4\const{P}^2}{(\lfloor n/L \rfloor)^2}\sum_{j=0}^{\lfloor n/L
    \rfloor -2} \sum_{\nu=1}^{\lfloor n/L \rfloor} \varrho^{\nu L} \nonumber\\
  & = & 4 \const{P}^2 \frac{\lfloor n/L \rfloor -1}{\lfloor n/L
    \rfloor} \frac{1}{\lfloor n/L \rfloor}  \sum_{\nu=1}^{\lfloor
      n/L \rfloor} \varrho^{\nu L}, \IEEEeqnarraynumspace \label{eq:typicalcov1}
\end{IEEEeqnarray}
and it follows from Ces\'aro's mean that this upper bound tends
to zero as $n$ tends to infinity. Likewise with \eqref{eq:typicaldelta2} we can upper bound the
second term on the RHS of \eqref{eq:typicalsum1} as
\begin{IEEEeqnarray}{lCl}
  \IEEEeqnarraymulticol{3}{l}{\frac{2}{(\lfloor n/L \rfloor)^2} \!\!\sum_{j=0}^{\lfloor n/L
    \rfloor -2} \sum_{\nu=1}^{\lfloor n/L \rfloor-1-j} 2\const{P}^2\sum_{\ell=1}^{j}\alpha_{\ell L}
    \alpha_{(\ell+\nu)L}\E{U_{\nu L+1}^2U_{1}^2}}\nonumber\\
  \qquad \qquad \quad & \leq & \frac{4 \const{P}^2}{(\lfloor n/L \rfloor)^2} \!\!\sum_{j=0}^{\lfloor n/L
    \rfloor -2} \sum_{\nu=1}^{\lfloor n/L \rfloor-1-j} \sum_{\ell=1}^{j}\alpha_{\ell L}
  \alpha_{(\ell+\nu)L}\E{U_{1}^4}\nonumber\\
  & < & 4 \const{P}^ 2\frac{\varrho^{2L}}{1-\varrho^{2L}} \E{U_{1}^4} \frac{\lfloor n/L \rfloor -1}{\lfloor n/L
    \rfloor}  \frac{1}{\lfloor n/L \rfloor} \sum_{\nu=1}^{\lfloor n/L
  \rfloor} \varrho^{\nu L}, \label{eq:typicalcov2}
\end{IEEEeqnarray}
where the first inequality follows from the Cauchy-Schwarz
inequality. As above, it follows from Ces\'aro's mean that this
upper bound tends to zero as $n$ tends to infinity.

It thus remains to show that the last term on the RHS of
\eqref{eq:typicalsum1} vanishes as $n$ tends to infinity. We have for
each $j=0,\ldots,\lfloor n/L \rfloor-2$
\begin{IEEEeqnarray}{lCl}
  \IEEEeqnarraymulticol{3}{l}{\sum_{\nu=1}^{\lfloor n/L \rfloor-1-j} \left(\sigma^2+\const{P}\sum_{\ell=1}^{j+\nu}\alpha_{\ell
        L}\right)\left(\sigma^2+\const{P}\sum_{\ell'=1}^{j}\alpha_{\ell'
        L}\right)\left(\E{U_{\nu L+1}^2U_{1}^2}-1\right)
  }\nonumber\\
  \qquad \quad & \leq & \sum_{\nu=1}^{\lfloor n/L \rfloor-1-j} \left(\sigma^2+\const{P}\sum_{\ell=1}^{j+\nu}\alpha_{\ell
      L}\right)\left(\sigma^2+\const{P}\sum_{\ell'=1}^{j}\alpha_{\ell'
      L}\right)\Bigl|\E{U_{\nu L+1}^2U_{1}^2}-1\Bigr|\nonumber\\
  & \leq & \sum_{\nu=1}^{\lfloor n/L \rfloor-1-j}
        \left(\sigma^2+\const{P}\alpha^{(L)}\right)^2 \Bigl|\E{U_{\nu
        L+1}^2U_{1}^2}-1\Bigr|\nonumber\\
  & \leq & \sum_{\nu=1}^{\lfloor n/L \rfloor}
        \left(\sigma^2+\const{P}\alpha^{(L)}\right)^2 \Bigl|\E{U_{\nu
        L+1}^2U_{1}^2}-1\Bigr|,
\end{IEEEeqnarray}
where the first inequality follows by upper bounding $\E{U_{\nu
    L+1}^2U_1^2}-1\leq\left|\E{U_{\nu L+1}^2 U_1^2}-1\right|$; and
    the second inequality follows by upper bounding $\sum_{\ell=1}^j
    \alpha_{\ell L} \leq \sum_{\ell=1}^{j+\nu}\alpha_{\ell
    L}\leq\sum_{\ell=1}^{\infty}\alpha_{\ell L}=\alpha^{(L)}$.
The last term on the RHS of \eqref{eq:typicalsum1} is therefore
upper bounded by
\begin{IEEEeqnarray}{lCl}
  \IEEEeqnarraymulticol{3}{l}{\frac{2}{(\lfloor n/L \rfloor)^2} \sum_{j=0}^{\lfloor n/L
      \rfloor -2} \sum_{\nu=1}^{\lfloor n/L \rfloor-1-j}\left(\sigma^2+\const{P}\sum_{\ell=1}^{j+\nu}\alpha_{\ell
        L}\right)\left(\sigma^2+\const{P}\sum_{\ell'=1}^{j}\alpha_{\ell'
        L}\right)\left(\E{U_{\nu L+1}^2U_{1}^2}-1\right)}\nonumber\\
  \qquad \qquad \quad & \leq & \frac{2}{(\lfloor n/L \rfloor)^2} \sum_{j=0}^{\lfloor n/L
    \rfloor -2} \sum_{\nu=1}^{\lfloor n/L \rfloor}
  \left(\sigma^2+\const{P}\alpha^{(L)}\right)^2 \Bigl|\E{U_{\nu
      L+1}^2U_{1}^2}-1\Bigr|\nonumber\\
  & = & 2  \left(\sigma^2+\const{P}\alpha^{(L)}\right)^2 \frac{\lfloor n/L \rfloor -1}{\lfloor n/L
    \rfloor}  \frac{1}{\lfloor n/L \rfloor} \sum_{\nu=1}^{\lfloor n/L
    \rfloor} \Bigl|\E{U_{\nu L+1}^2U_{1}^2}-1\Bigr|. \label{eq:typicalcov3}
\end{IEEEeqnarray}
It follows now from the weakly-mixing property of $\{U_k\}$ that
\cite[Thm.~6.1]{petersen83}
\begin{equation*}
  \lim_{n \to \infty} \frac{1}{\lfloor n/L \rfloor} \sum_{\nu=1}^{\lfloor n/L
  \rfloor} \Bigl|\E{U_{\nu L+1}^2U_{1}^2}-1\Bigr| = \lim_{n\to\infty} \frac{1}{\lfloor n/L \rfloor} \sum_{\nu=1}^{\lfloor n/L
  \rfloor} \Bigl|\E{U_{\nu L+1}^2U_{1}^2}-\E{U_{\nu L+1}^2}\E{U_1^2}\Bigr|= 0
\end{equation*}
so that the last term on the RHS of \eqref{eq:typicalsum1}
vanishes as $n$ tends to infinity.

Thus \eqref{eq:typicalcov3}, \eqref{eq:typicalcov2}, and
\eqref{eq:typicalcov1} show that \eqref{eq:typicalsum1}
vanishes as $n$ tends to infinity which in turn shows, along with
\eqref{eq:typicalvar} and \eqref{eq:typicalfinite}, that
\begin{IEEEeqnarray*}{c}
\lim_{n \to \infty} \Var{\frac{1}{\lfloor n/L \rfloor} \|\vect{Y}\|^2} = 0.
\end{IEEEeqnarray*}
Together with \eqref{eq:typicalcheby1}, this proves
\eqref{eq:typicalP1}. The proof of \eqref{eq:typicalP2} follows along
the same lines.


\end{document}